\documentclass[structabstract]{aa}
\usepackage{graphicx}
\usepackage{natbib}
\bibpunct{(}{)}{;}{a}{}{,} 
\usepackage{amsmath}
\usepackage{txfonts}
\usepackage{rotating}
\begin{document}
   \title{The dynamics and star-forming potential of the massive Galactic centre cloud G0.253+0.016\thanks{The FITS images presented in Figs. \ref{SMAcontinuum}, \ref{combined_1.3mm}, \ref{compare_line_cont} and \ref{IRAM13COfig} are available in electronic form at the CDS via anonymous ftp to cdsarc.u-strasbg.fr (130.79.128.5) or via http://cdsarc.u-strasbg.fr/viz-bin/qcat?J/A+A/568/A56}}

   \author{K. G. Johnston \inst{1}
   \and H. Beuther \inst{1}
   \and H. Linz \inst{1}
   \and A. Schmiedeke \inst{2}
   \and S. E. Ragan \inst{1}
   \and Th. Henning \inst{1} }

   \institute{Max Planck Institute for Astronomy, K\"onigstuhl 17, D-69117 Heidelberg, Germany \\
              \email{johnston@mpia.de}
        \and       
         I. Physikalisches Institut der Universit\"at zu K\"oln, Z\"ulpicher Strasse 77, 50937 K\"oln, Germany
           }

   \date{Received 3 April 2014, Accepted 3 June 2014}

  \abstract{The massive infrared dark cloud G0.253+0.016 projected $\sim$45\,pc from the Galactic centre contains $\sim$10$^5$\,M$_{\odot}$ of dense gas whilst being mostly devoid of observed star-formation tracers.}
  {Our goals are therefore to scrutinise the physical properties, dynamics and structure of this cloud with reference to its star-forming potential.}
  {We have carried out a concerted SMA and IRAM 30m study of this enigmatic cloud in dust continuum, CO isotopologues, several shock tracing molecules, as well as H$_2$CO to trace the gas temperature. In addition, we include ancillary far-IR and sub-mm \textit{Herschel} and SCUBA data in our analysis.}
  {We detect and characterise a total of 36 dust cores within G0.253+0.016 at 1.3\,mm and 1.37\,mm, with masses between 25 and approximately 250\,M$_{\odot}$, and find that the kinetic temperature of the gas traced by H$_2$CO ratios is $>$320\,K on size-scales of $\sim$0.15\,pc. Analysis of the position-velocity diagrams of our observed lines shows broad linewidths and strong shock emission in the south of the cloud, indicating that G0.253+0.016 is colliding with another cloud at v$_{_{\rm LSR}}\sim 70$\,km\,s$^{-1}$. We confirm via an analysis of the observed dynamics in the Central Molecular Zone that it is an elongated structure, orientated with Sgr\,B2 closer to the Sun than Sgr\,A*, however our results suggest that the actual geometry may be more complex than an elliptical ring. We find that the column density Probability Distribution Function (PDF) of G0.253+0.016 derived from SMA and SCUBA dust continuum emission is log-normal with no discernible power-law tail, consistent with little star formation, and that its width can be explained in the framework of theory predicting the density structure of clouds created by supersonic, magnetised turbulence. We also present the $\Delta$-variance spectrum of this region, a proxy for the density power spectrum of the cloud, and show it is consistent with that expected for clouds with no current star formation. Finally, we show that even after determining a scaled column density threshold for star formation by incorporating the effects of the increased turbulence in the cloud, we would still expect ten stars with masses $>$15\,M$_{\odot}$ to form in G0.253+0.016. If these cannot be accounted for by new radio continuum observations, then further physical aspects may be important, such as the background column density level, which would turn an absolute column density threshold for star formation into a critical over-density.}
{We conclude that G0.253+0.016 contains high-temperatures and wide-spread shocks, displaying evidence of interaction with a nearby cloud which we identify at v$_{_{\rm LSR}}\sim 70$\,km\,s$^{-1}$. Our analysis of the structure of the cloud can be well-explained by theory of magnetised turbulence, and is consistent with little or no current star formation. Using G0.253+0.016 as a test-bed of the conditions required for star formation in a different physical environment to that of nearby clouds, we also conclude that there is not one column density threshold for star formation, but instead this value is dependant on the local physical conditions.}
    
 
\titlerunning{The massive Galactic centre cloud G0.253+0.016}
\authorrunning{K. G. Johnston et al.}
 
\maketitle
 
\section{Introduction \label{intro}}

Determining how massive clusters ($10^{3} - 10^{5}$\,M$_{\odot}$)
form has a profound effect on
how we interpret observations of star formation in external galaxies.
As the majority of stars form in clusters \citep{lada03,de-wit05}, and because massive
clusters yield -- either via statistics or by virtue of their physical
conditions -- the most massive stars, these clusters are the engines
which produce the objects that dominate the luminosity of galaxies.
Therefore, uncovering how and where these massive clusters can form is
of central importance in understanding how galaxies evolve, and may
provide hints as to how cluster formation proceeds at all masses.

To understand the conditions that lead to cluster formation, it is
necessary to observe the structure of the gas and dust, as well as
kinematics, of a cluster forming cloud before star formation processes
begin blurring the initial conditions. Observing the formation of
massive clusters within our own Galaxy has the obvious advantage of
resolving clouds that typically fall within a single resolution element for observations
of distant galaxies. 

One of the most exceptional candidates for a massive
cluster progenitor is the cloud G0.253+0.016 near the Galactic centre (e.g., \citealt{lis94,lis98,longmore12}).
The global dust properties of G0.253+0.016 (also known as M0.25+0.01) have been shown by previous authors \citep[e.g.][]{lis94,lis98,lis01,longmore12, immer12} to be cold
($\sim$18-27\,K), dense (n$\sim$7.3\,$\times$10$^{4}$ - 6\,$\times$10$^{5}$ cm$^{-3}$)
and massive (M=1.3 - 7 $\times10^{5}$\,M$_{\odot}$).
However, minimal evidence has been found for ongoing star formation. The H$_2$O maser
and 8.4\,GHz radio continuum observations of \citet{lis94} uncovered a
single H$_2$O maser near the 350\,$\mu$m continuum peak position, but
no coincident radio emission. \textit{Spitzer} and \textit{Herschel} observations have also confirmed that no
protostars are seen in the cloud in the mid- or far-IR up to 70\,$\mu$m \citep[e.g.,][]{longmore12}. 
\citet{rodriguez13} presented Very Large Array (VLA) 1.3 and 5.6\,cm radio continuum observations
of G0.253+0.016 which detected three compact radio sources towards the eastern 
edge of the cloud, which have thermal spectral indices and correspond to B0.5 stars. 
These could be signposts of massive star formation in G0.253+0.016; however, as will be 
discussed below, they do not correspond to dense cores of gas traced by mm emission. 

It is interesting to note that the similarly massive cloud Sgr B2, which is
not far away from G0.253+0.016, actively forms stars and is one of the
most prominent sites of star formation in the Galaxy. Therefore, how can such a 
massive and dense molecular cloud as G0.253+0.016 currently not be forming stars? 
A possible explanation for the high star-formation activity in Sgr B2 is that one of the 
dust lanes associated with the Galactic bar 
intersects with the $x_{2}$ orbits of the gas in the Central Molecular Zone (CMZ)
at the position of Sgr B2, and thus we would expect an elevated level of star 
formation at this position. Yet this does not straightforwardly explain the comparatively
lesser degree of star formation towards the other point of intersection with the 
dust lanes of the bar on the opposite side of the CMZ, Sgr C \citep[e.g.,][]{kendrew13}. 
One possible explanation could involve an asymmetry in the material falling inwards 
along the two dust lanes of the bar, or in the gas orbiting the CMZ, 
providing less material for collision at the position of Sgr C. Alternatively, the star 
formation in Sgr B2 could be enhanced by its recent passage close 
to the Galaxy's central black hole, Sgr\,A* \citep{longmore13}. Of course, the observed differences between
Sgr B2 and Sgr C may be due to a combination of these effects.

Recently, G0.253+0.016 has also been studied using MALT90 and APEX observations \citep{rathborne14}, and ALMA SO observations have shown evidence for a cloud-cloud collision with G0.253+0.016 \citep{higuchi14}. In fact, such cloud-cloud collisions may provide a way to collect enough dense gas to produce massive clusters \citep[e.g.,][]{fukui14}.

With reference to these possible scenarios, in this work we aim to determine the 
current state of star formation in G0.253+0.016, as well as its star-forming fate 
(i.e., whether it will form a star cluster), by investigating the structure of the cloud, 
its internal dynamics, its interaction with the CMZ environment and its local stability to collapse.

Section \ref{obs} outlines our 1.3\,mm Submillimeter Array (SMA) and Institut de Radioastronomie Millim\'{e}trique (IRAM) 30m observations, as well as ancillary far-IR and sub-mm \textit{Herschel} and Submillimeter Common-User Bolometer Array \citep[SCUBA;][]{holland99} data. Section \ref{result} presents our observational results from the continuum and line observations, including the column density Probability Distribution Function (PDF) of G0.253+0.016, and determination of temperatures from H$_2$CO. Section \ref{discussion} presents our discussion, which covers the topics of the internal dynamics of G0.253+0.016, the interaction of G0.253+0.016 with its environment, as well as its current star-formation activity and potential. Our conclusions are given in Section~\ref{conclusions}.

\section{Observations and Archival Data}
\label{obs}

\subsection{SMA Line and Continuum Observations}
Observations of G0.253+0.016 were conducted on 9 June 2012 with the SMA in its compact array configuration under good weather conditions (the optical depth at 225\,GHz was below 0.1). The two 4\,GHz sidebands were placed at 218.9 and 230.9\,GHz (1.37 and 1.30\,mm), with each of the 48 SMA correlator chunks per sideband having 128 channels and a spectral resolution of 0.812\,MHz or 1.1\,kms$^{-1}$. Seven of the eight SMA antennas were available for the observation, providing baseline lengths between 16.4\,m and 77.0\,m and thus a largest angular scale of approximately 20-21$''$. Table \ref{obstable} lists the observed lines and continuum, their frequencies, synthesised beam sizes and the rms noise in the final images. The observations consisted of a 6-pointing mosaic covering G0.253+0.016, with spacing between pointings of approximately half the primary-beam width (54-58$''$). The total time on-source was 4.5\,hr. The gain, bandpass and flux calibrators were 1733-130, 3C279 and Titan. Data reduction was carried out using the MIR package\footnote{the MIR package and cookbook can be found at https://www.cfa.harvard.edu/~cqi/mircook.html}, followed by imaging in MIRIAD \citep{sault95}. 

\begin{table*}[]
\caption{Summary of observed/imaged lines}
\centering
\begin{tabular}{llllll}
\hline
Facility & Line or Continuum & Frequency & Angular Resolution & Imaged Spectral & Sensitivity \\
 & & (GHz) & (arcseconds) & Resolution (kms$^{-1}$) & (mJy\,beam$^{-1}$) \\
\hline
SMA & SiO (5-4)                     & 217.105         & 4.4$\times$2.9, PA 0.7$^{\circ}$ & 2, 5 & 68, 48 \\
SMA & H$_2$CO 3(0,3) - 2(0,2)  & 218.222    & 4.3$\times$2.9, PA -1.0$^{\circ}$ & 2, 5 & 70, 50 \\
SMA & CH$_{3}$OH 4(2,2) - 3(1,2)-E & 218.440 & 4.3$\times$2.9, PA -1.1$^{\circ}$ & 2, 5 & 76, 56 \\
SMA & H$_2$CO 3(2,2) - 2(2,1)  & 218.476 \tablefootmark{a} &  &  \\
SMA & H$_2$CO 3(2,1) - 2(2,0)  & 218.760     & 4.3$\times$2.9, PA -1.0$^{\circ}$ & 2, 5 & 65, 43\\
SMA & C$^{18}$O (2-1)               & 219.560     & 4.3$\times$2.9, PA 1.0$^{\circ}$ & 2, 5 & 74, 54 \\
SMA & HNCO 10(0,10) - 9(0, 9) &  219.798      & 4.3$\times$2.9, PA 0.7$^{\circ}$ & 2, 5 & 67, 46 \\
SMA & SO 6(5) - 5(4)                   & 219.949 & 4.3$\times$2.9, PA 1.0$^{\circ}$ & 2, 5 & 68, 47 \\
SMA & $^{13}$CO (2-1)               & 220.399      & 4.2$\times$2.9, PA 0.7$^{\circ}$ & 2, 5&  240, 220 \\
SMA & CH$_{3}$OH 8(-1, 8) - 7(0,7)-E & 229.759  & 4.3$\times$2.7, PA 3.6$^{\circ}$ & 2, 5 & 74, 51 \\
SMA & $^{12}$CO (2-1)               & 230.538     & 4.3$\times$2.7, PA 4.4$^{\circ}$ & 2, 5 & 890, 890 \\
SMA & 4\,GHz continuum            & 216.9 - 220.9        & 4.3$\times$2.9, PA -1.1$^{\circ}$ &  & 2.5 \\
SMA & 4\,GHz continuum            & 228.9 - 232.9       & 4.3$\times$2.7, PA 4.0$^{\circ}$ & & 2.5 \\
IRAM 30m & $^{13}$CO (2-1)               & 220.399      & 11.8 & 2, 5 & 2500, 1600  \\
IRAM 30m & $^{12}$CO (2-1)               & 230.538     & 11.2 & 2, 5 & 6000, 5000  \\
SMA + IRAM 30m & $^{13}$CO (2-1)                  & 220.399      & 4.2$\times$2.9, PA 0.7$^{\circ}$ & 2, 5 &  200, 140 \\
SMA + IRAM 30m & $^{12}$CO (2-1)                   & 230.538     & 4.3$\times$2.7, PA 4.4$^{\circ}$ & 2, 5 &  500, 500 \\
\end{tabular}
\tablefoot{
\tablefoottext{a}{Blended with CH$_{3}$OH 4(2,2)- 3(1,2)-E}
}
\label{obstable}
\end{table*}

\subsection{IRAM 30m Line Observations}
\label{obsmm} 

Observations were also carried out on 16 and 22 October 2012 using the Eight MIxer Receiver (EMIR) installed on the IRAM 30m telescope on Pico Veleta, Spain. We observed with the Fourier Transform Spectrometer (FTS) backend in the E2 band which has 16\,GHz bandwidth per polarisation. The two 8\,GHz bandwidth sidebands were placed at 217.3 and 233.0\,GHz, so that they covered the same frequencies as the SMA observations. The spectra had a resolution of 0.2\,MHz or approximately 0.3\,kms$^{-1}$. An area of 3$' \times 4'$ (in R.A. and Dec. respectively) covering G0.253+0.016 was mapped in three sub-fields using the on-the-fly position switching mode. Five on-the-fly maps were performed by covering three sub-fields, scanning first in right ascension, then declination. Unfortunately, due to poor weather conditions, we did not obtain the data for the sixth map which should have been scanned in declination over the upper 3$' \times 80''$ of the map, decreasing the sensitivity in this area by $\sqrt{2}$. The total time on-source was 77.5\,min. Spectra and maps were made using the GILDAS package. When averaging over the entire map, all lines detected in the SMA observations were detected with the IRAM 30m. In addition, several weak unidentified lines at 217.823, 219.875, 229.900 and 229.931\,GHz were also detected in the average spectrum over the map. The $^{12}$CO and $^{13}$CO line emission had sufficient signal-to-noise to create maps and to be combined with the SMA line channel maps. The combination was carried out using the feather task in CASA. The beam size and rms noise for the $^{12}$CO and $^{13}$CO uncombined IRAM 30m and combined SMA + IRAM 30m maps are given in Table \ref{obstable}.

\subsection{Herschel far-IR and SCUBA 450\,$\mu$m data}
\label{SD} 

We also made use of archival data from the {\it Herschel} satellite \citep{pilbratt10}.  Photometric data at 70, 160, 250, 350 and 500\,$\mu$m were obtained with the PACS \citep{poglitsch10} and SPIRE \citep{griffin10} bolometric cameras on September 7, 2010 (operational day 481) as part of the Hi-Gal survey of the Galactic plane \citep{molinari10a}. These observations were performed in parallel-mode (PACS and SPIRE in tandem) with a fast scan rate of 60$''$s$^{-1}$, and the data set used (obsids 1342204102, 1342204103) covers a $2^\circ \times 2^\circ$ patch of the Galactic plane in two perpendicular scan directions, centred on the Galactic coordinates [0.0, 0.0] and including G0.253+0.016. For PACS, we downloaded the relevant Level~1 data from the \textit{Herschel} Science Archive \citep{leon09}, which were produced by the HCSS 10.3.0 bulk reprocessing of the raw data. We used the HIPE software \citep[track 12.0, build 2491][]{ott10} to apply an additional correction of the bolometer response, based on the drift of the evaporator temperature \citep{balog13} before using the SCANAMORPHOS software \citep{roussel13}, version 22.0, for further drift corrections, 1/$f$ noise removal, and final mapping. For SPIRE, we downloaded the Level~2 data (calibrated for extended emission) from the HCSS 10.3.0 bulk reprocessing and combined the two scan directions with the HIPE mosaic task.  The pixel sizes for the PACS and SPIRE data, listed by increasing wavelength, are 3.2, 3.2, 6, 10 and 14$''$ and the assumed FWHM beam sizes are 8.8, 13.0, 23.3, 30.3 and 42.4$''$. The PACS 70 and 160\,$\mu$m beam was derived from 2D gaussian fits to point sources in the image, and the SPIRE beams were assumed to be the geometric mean of the values stated in Table~2 of \citet{Traficante11}. We refrained from using the convolution kernels published by \citet{aniano11} since they are not adapted to the specifics of the parallel-mode PSFs.

\begin{figure*}
\begin{center}
\sidecaption
\includegraphics[width=14cm]{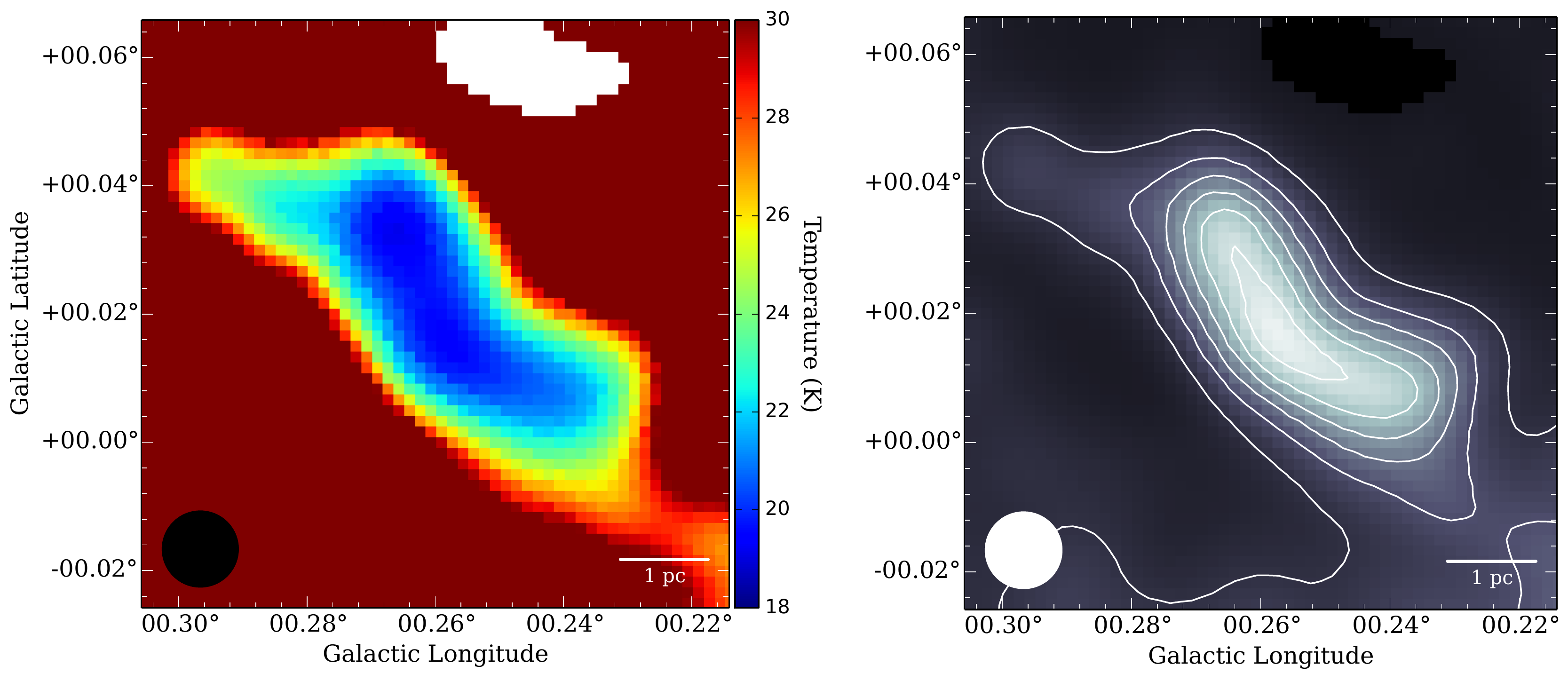}
\caption[]{\textit{Left:} Temperature map derived from SED fits to \textit{Herschel} images assuming $\beta = 1.75$. \textit{Right:} Model 450\,$\mu$m image derived from \textit{Herschel} data, assuming $\beta = 1.75$. Contours are 50, 100, 150 and 200, 250 and 300\,Jy\,beam$^{-1}$. Greyscale: -35 to 350 Jy\,beam$^{-1}$. The beam (FWHM 42.4$''$) is shown in the bottom left corner.}
\label{temp_map}
\end{center}
\end{figure*}

In addition, we obtained a 450\,$\mu$m SCUBA image covering G0.253+0.016 from the James Clark Maxwell Telescope Science Archive of the Canadian Astronomy Data Centre, under project code m98au64 \citep{pierce-price00}. The SCUBA 450\,$\mu$m beam consists of an 8$''$ FWHM inner beam and a 30$''$ first error beam with relative peaks of 0.94 and 0.06 \citep{hogerheijde00}. The maximum scale which the observations are sensitive to is determined by several times the maximum chop throw, which was 65$''$. This was estimated to be $\sim$10\,pc or 4$'$ by \citet{pierce-price00}, and thus most of the emission from G0.253+0.016 should be recovered. Flux calibration of the 450\,$\mu$m SCUBA image was performed by comparing a model of the 450\,$\mu$m emission derived from the \textit{Herschel} data. We determined the 450\,$\mu$m model image by modelling the spatially varying emission of G0.253+0.016 by fitting modified blackbodies (blackbodies multiplied by a wavelength-dependent opacity) to each pixel of the \textit{Herschel} 160-500\,$\mu$m PACS and SPIRE images. Firstly, all the three \textit{Herschel} images at shorter wavelengths were convolved to a resolution of 42.4$''$, and reprojected to the same FITS header using Montage \citep{jacob09}. The large-scale background was then removed from each image using the algorithm based on that outlined in \citet{battersby11}, which we now describe. When creating the model of the background emission, we smoothed each image by 20$'$ and used this image directly instead of fitting gaussians in latitude to the emission. A source mask was created by masking emission above 3\,$\sigma$ in a difference image created by subtracting the smoothed 500\,$\mu$m image from the original 500\,$\mu$m image. This process was repeated, using the mask derived in the previous iteration until the 3\,$\sigma$ level of the masked difference image converged. This source mask was applied to each image before it was smoothed to determine the background to be subtracted. When fitting the modified blackbodies, we applied temperature and $\beta$ (the opacity power-law index) dependent colour corrections to the PACS 100\,$\mu$m model images.\footnote{Section 4.3 of http://herschel.esac.esa.int/twiki/pub/Public/ \\ PacsCalibrationWeb/cc\_report\_v1.pdf} We did not apply any colour corrections to the SPIRE models as these were small: less than 4\,\% for temperatures between 15 and 40\,K \citep{griffin13}. The results of the fit included: the flux scaling of the fit, which essentially gives the column density, and the temperature at each point in the image. We produced three model images by evaluating the model at 450\,$\mu$m when assuming three fixed values of $\beta$: 1.5, 1.75 and 2.0. Figure \ref{temp_map} presents the temperature map and model 450\,$\mu$m image for $\beta = 1.75$. The derived $\beta = 1.75$ temperature map is in good agreement ($<$10\%) with the map presented by \citet{longmore12}; the temperatures across G0.253+0.016 range from $\sim$19 to $\sim$30\,K. There is a remaining background contribution from the CMZ after large-scale background subtraction in the temperature map presented here and in \citet{longmore12}, which will slightly increase the measured temperature of the cloud. If a higher threshold than 3\,$\sigma$ was used to create the background mask, over-subtraction of the source occurred at the edges of the cloud, leading to incorrect temperature and flux values at these positions. Therefore we opted to increase the size of the mask to 3\,$\sigma$ to obtain correct relative values of temperature and flux across G0.253+0.016, while noting that the absolute temperature may be overestimated at most by 10\% or 3\,K within the region covered by our SMA observations. 

After the model 450\,$\mu$m image was produced, the SCUBA 450\,$\mu$m image was convolved to the same resolution (42.4$''$), using a similar method to that of \citet{aniano11}. This consisted of taking the inverse transform of the ratio of Fourier transforms of the SPIRE 500\,$\mu$m and SCUBA beams. We made a model of the SCUBA 450$\mu$m beam using the parameters of the first two components of the beam given in Table 2 of \citet{hogerheijde00}, consisting of an 8$''$ FWHM inner beam and a 30$''$ first error beam with relative peaks of 0.94 and 0.06, with a corresponding effective beam size of 10.7$''$. The images were then reprojected to the same FITS header and the relationship between the pixel values was fit by a slope and offset. The offset corresponds to the remaining background in the 450\,$\mu$m model image, and the slope corresponds to the calibration factor, which were found to be [6.053, 5.619, 4.993] Jy\,beam$^{-1}$ and [0.65, 0.62, 0.59] respectively for $\beta = $[1.5, 1.75, 2.0]. Thus the calibration factor of the original SCUBA 450\,$\mu$m image to the \textit{Herschel} calibration is 0.62$\pm$0.03, with the error determined from the range in assumed values of $\beta$.

As combining the two SMA sideband continuum images did not significantly increase the continuum sensitivity and would not straightforwardly represent one frequency, we decided to combine the SMA 1.3\,mm upper sideband continuum image with the single dish SCUBA image. To do this we scaled our calibrated SCUBA 450\,$\mu$m image to 1.3\,mm, by evaluating the model of the emission derived from the \textit{Herschel} data at 1.3\,mm and finding the pixel-to-pixel ratio between the 450\,$\mu$m and 1.3\,mm model images. We used this ratio image to then scale the flux calibrated SCUBA 450\,$\mu$m image to 1.3\,mm. This resulted in three images at 1.3\,mm, one for each of the assumed values of~$\beta$. These three images were then combined with the SMA 1.3\,mm image using the feather task in CASA.

\section{Results}
\label{result}

\subsection{SMA Continuum Emission}
\label{SMA_cont}

\begin{figure*}
\begin{center}
\includegraphics[width=14cm]{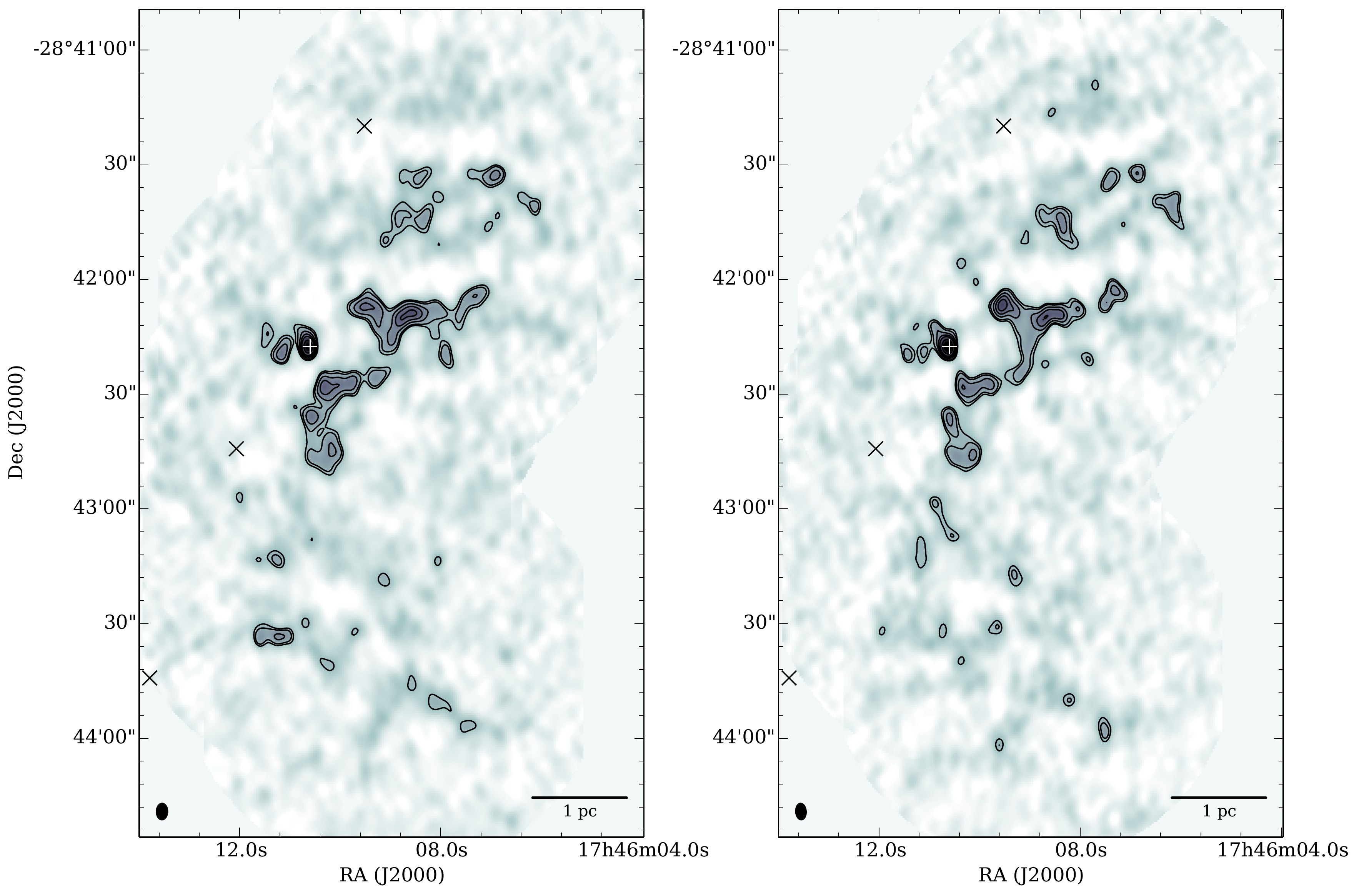}
\caption{\textit{Left panel:} Map of the 218.9\,GHz or 1.37\,mm continuum emission observed with the SMA. Contours are -5, 5, 6, 8, 10, 12, 16 and 20 $\times$ rms noise = 2.5\,mJy\,beam$^{-1}$. Greyscale: -2 to 50 mJy\,beam$^{-1}$. The synthesised beam is shown in the bottom left corner: 4.3$''$ $\times$ 2.9$''$, P.A. = -1.1$^{\circ}$. \textit{Right panel:} Map of the 230.9\,GHz or 1.30\,mm continuum emission observed with the SMA. The contours and stretch are the same as in the left panel. The synthesised beam is shown in the bottom left corner: 4.3$''$ $\times$ 2.7$''$, P.A. = 4.0$^{\circ}$. In both panels the plus sign marks the position of the water maser reported by \citet{lis94} and the crosses mark (respectively from north to south) the positions of the 1.3\,cm sources VLA 4 to 6 from \citet{rodriguez13}.}
 \label{SMAcontinuum}
 \end{center}
\end{figure*}

The two panels of Fig. \ref{SMAcontinuum} show respectively the lower and upper sideband continuum observed with only the SMA, centred at 230.9 and 218.9\,GHz, or 1.30 and 1.37\,mm. The position of the maser observed by \citet{lis94} is marked as a plus sign, and the three crosses mark the positions of the 20.9\,GHz or 1.3\,cm sources VLA 4 to 6 from \citet{rodriguez13}. The maser coincides with the brightest 1\,mm continuum source in the region, however no obvious dust emission is associated with the \citet{rodriguez13} 1.3\,cm sources. The brightest 1\,mm continuum source detected in our observations is also the same as the 1.1\,mm source detected by \citet{kauffmann13} at 280\,GHz, however further cloud structure is also apparent above 5\,$\sigma$ in our 1.30 and 1.37\,mm maps.

To characterise the continuum emission detected at 1.30 and 1.37\,mm, we produced dendrograms \citep[e.g.][]{rosolowsky08, goodman09} using the \textit{astrodendro} software\footnote{http://www.dendrograms.org}. We chose the flux limit above which fluxes were measured to be 2\,$\sigma$; the required peak flux of the leaves of the dendrogram, i.e. the highest density structures which had no further structures embedded within them, to be at least 5\,$\sigma$; and the required flux difference between embedded structures $\Delta F$ to be $\sigma$. The \textit{astrodendro} code was run separately on both the 1.37 and 1.30\,mm (lower and upper side bands, respectively) continuum images. The leaves are listed in Table~\ref{dendro_contin_table}, where we refer to them as cores. Confirmed cores which were detected in both images and had an overlap in pixels greater than 30 pixels are given first, listing the core number, peak position, peak flux $S_{\rm peak}$, integrated flux $S_{\rm int}$, mass $M$, peak column density $N_{peak}$ and geometric mean diameter $D$ for the 1.37 and 1.30\,mm images (or lower and upper sidebands) respectively. 

\begin{figure}
\begin{center}
\includegraphics[width=8cm]{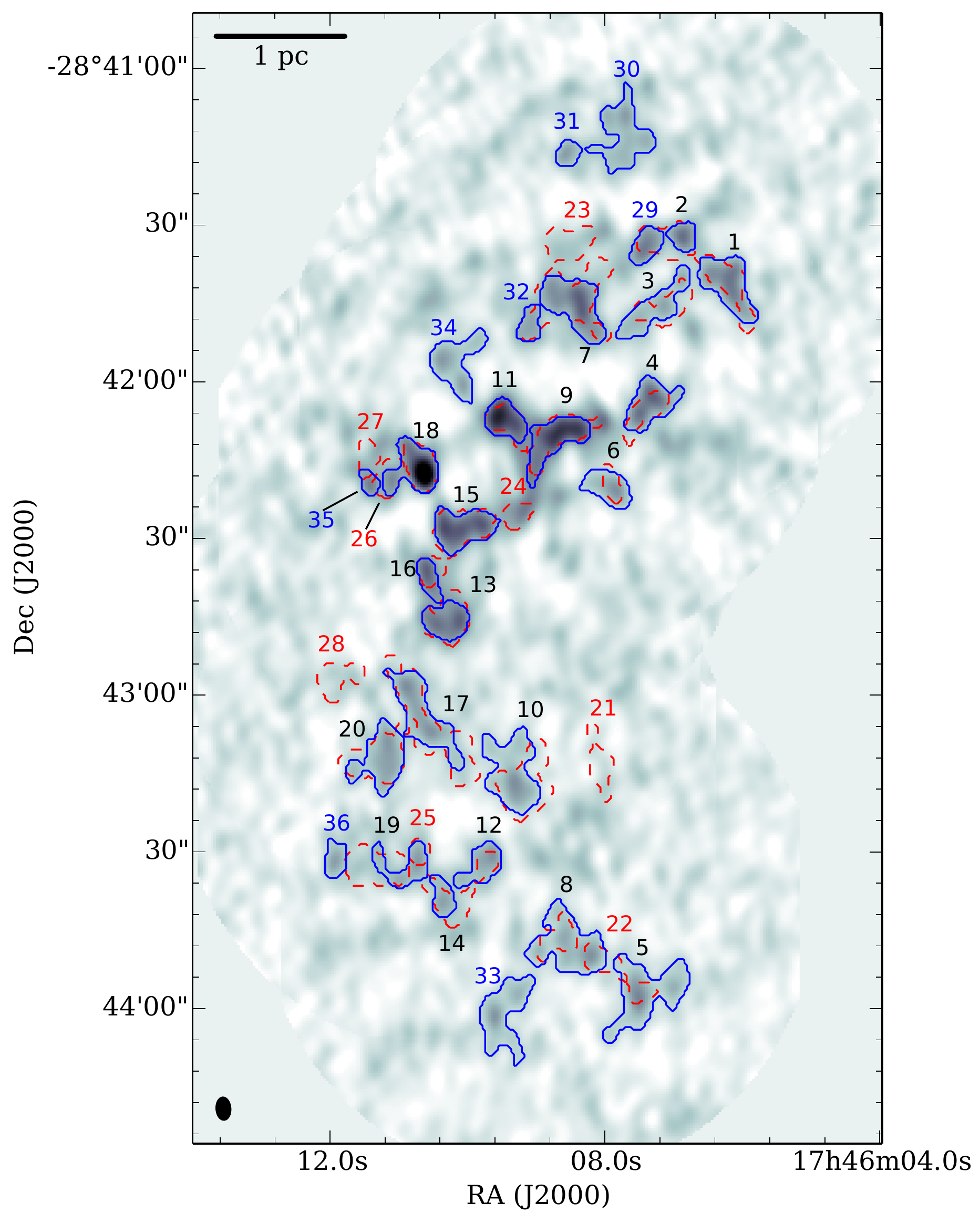}
\caption{Map of the cores detected in 218.9 and 230.9 GHz (1.37mm and 1.30mm) continuum emission observed with the SMA, in red dashed and blue solid contours respectively. The 1.30\,mm continuum emission is shown in greyscale, ranging between -2.5 and 40 mJy\,beam$^{-1}$. The cores are numbered as listed in Table~\ref{dendro_contin_table}. The synthesised beam is shown in the bottom left corner: 4.3$''$ $\times$ 2.7$''$,~P.A.=4.0$^{\circ}$.}
 \label{SMAcontinuum_numbered}
 \end{center}
\end{figure}

The remaining cores, which were detected in one band only or overlapped with cores detected in the other image to a lesser degree, are listed at the bottom of Table~\ref{dendro_contin_table}. Figure~\ref{SMAcontinuum_numbered} shows an overlay in contours of the sources detected in the two continuum images, with the allocated core numbers shown next to each core. It is probable that the emission from cores only detected in one band is real, because given a gaussian distribution of noise, emission which is intrinsically at the level of 5$\sigma$ will be detected above this in half of cases, whereas the chance that a pixel in the images which has no intrinsic emission is boosted above 5$\sigma$ by the noise is less than once for a million-pixel image (based on simulations of images with injected noise measured with the \textit{astrodendro} software). As there are 9.1$\times 10{^4}$ pixels covered by the mosaic pattern in each continuum image, we would expect less than one false detection every ten images; given two images, there is thus a $<$20\% chance that there is one false source in one of them. Therefore, it is likely that the majority, if not all, of the nine cores detected without a counterpart in the other continuum image are real. These nine cores consist of the 16 cores listed without a match in Table~\ref{dendro_contin_table}, minus the seven cores which were partially associated with other cores (marked \textit{a} in the table).

The masses of the cores were determined using the equation,

\begin{equation} M = \frac{g S_{\rm int} d^2}{B(\nu,T) \kappa_{\nu}} \end{equation}

\noindent where $g$ is the gas-to-dust ratio, $d$ is the distance, $B(\nu,T)$ is the black body function, which is a function of frequency $\nu$ and temperature $T$, and $\kappa_{\nu}$ is the frequency-dependent opacity. We assumed a distance of 8.4$\pm$0.6\,kpc \citep{reid09}, and used the temperature maps derived from the \textit{Herschel} data in Section \ref{SD} to determine the temperature at the core positions. The gas-to-dust ratio for the solar neighbourhood was calculated to be 154, given the ratio between the mass of dust and mass of hydrogen of 0.0091 \citep{draine11}. The gas-to-dust ratio in the Galactic centre is roughly half this value, due to a metallicity twice as high as the solar value \citep[e.g.][]{najarro09}. Thus we assumed a gas-to-dust ratio for the Galactic centre of 77. We assumed an opacity of 0.701\,cm$^{2}$g$^{-1}$ at 1.30\,mm, taken from \citet{ossenkopf94} for $n$=10$^{5}$\,cm$^{-3}$ and thin ice mantles. For the opacity at 1.37\,mm, we fit a line in log space to the same opacities above wavelengths of 40\,$\mu$m, where the dependence of opacity on wavelength becomes a power law, and extrapolated the 1.3\,mm opacity to 1.37\,mm to obtain 0.602\,cm$^{2}$g$^{-1}$. These opacities are uncertain by a factor of two or less \citep{ossenkopf94} and thus dominate the overall mass uncertainty. The peak column densities were calculated in the same fashion, instead using the peak fluxes and dividing by the relevant SMA beam area. The geometric mean diameter was calculated from the area covered by each dendrogram core structure.

Continuum observations carried out with the VLA at 24.1 to 36.4\,GHz (1.2 to 0.8\,cm) with $\sim2''$ resolution display emission towards 20 of the detected 1.3\,mm cores (Mills et al., in preparation). Those which are associated with 24.8\,GHz or 1.2\,cm emission above 0.125\,mJy\,beam$^{-1}$ in a 2.2$''$ $\times$ 1.9$''$ beam (Mills et al., in preparation) are marked with a Y in the final column of Table~\ref{dendro_contin_table}. The brightest 1.2\,cm continuum source that is coincident with a core (Core 9), has a peak flux density of 0.55\,mJy\,beam$^{-1}$ at 1.2\,cm and a flat spectral index, indicating optically thin free-free emission. Core 9 has a peak flux of 30.3\,mJy\,beam$^{-1}$ at 1.3\,mm, therefore the extrapolated ionised gas emission should only contribute a few percent to the flux of this source at 1.3\,mm. The most obvious coincidences of dust and 1.3\,cm ionized gas emission exist for Cores 9, 13 and 26 (also the edge of core 18). Thus these may be indicating the presence of high-mass star formation within G0.253+0.016. 

\begin{table*}
\caption{Measured and calculated properties of the cores detected in SMA continuum, exclusive of the combination with the short-spacing SCUBA emission presented in Section \ref{SMA_SCUBA_result}.}  
\resizebox{\linewidth}{!}{%
\begin{tabular}{llllllllllllll}
\hline
Core & \multicolumn{6}{l}{Lower Side Band} & \multicolumn{6}{l}{Upper Side Band} &\\
number & Peak position & S$_{\rm peak}$ & S$_{\rm int}$ & Mass & Column density & Diameter & Peak position & S$_{\rm peak}$ & S$_{\rm int}$ & Mass & Column density & Diameter & 24.8\,GHz \\
 & (J2000) & (mJy\,beam$^{-1}$) & (mJy) & (M${_\odot}$) & (g\,cm$^{-2}$) & (pc)\tablefootmark{b} & (J2000) & (mJy\,beam$^{-1}$) & (mJy) & (M${_\odot}$) & (g\,cm$^{-2}$) & (pc)\tablefootmark{b} & emission\tablefootmark{c} \\
\hline
 1 &  17:46:06.1 -28:41:41.1 &     23.1 &     63.2 &      119 &   0.39 &   0.35 & 17:46:06.2 -28:41:40.6 &     26.3 &    101.7 &      151 &   0.37 &   0.41 &  N \\ 
 2 &  17:46:06.9 -28:41:32.6 &     22.7 &     60.9 &      123 &   0.41 &   0.36 & 17:46:06.9 -28:41:32.6 &     20.9 &     24.6 &       39 &   0.32 &   0.22 &  N \\ 
 3 &  17:46:07.1 -28:41:46.1 &     13.5 &     37.2 &       74 &   0.24 &   0.33 & 17:46:07.1 -28:41:45.6 &     12.8 &     44.3 &       69 &   0.19 &   0.38 &  N \\ 
 4 &  17:46:07.3 -28:42:04.6 &     22.5 &     50.9 &       95 &   0.37 &   0.29 & 17:46:07.3 -28:42:03.1 &     25.7 &     76.3 &      112 &   0.36 &   0.37 &  Y \\ 
 5 &  17:46:07.5 -28:43:56.6 &     28.3 &     29.1 &       44 &   0.38 &   0.19 & 17:46:07.5 -28:43:58.1 &     36.8 &    158.7 &      185 &   0.41 &   0.51 &  N \\ 
 6 &  17:46:07.9 -28:42:19.6 &     22.1 &     25.6 &       47 &   0.36 &   0.21 & 17:46:07.9 -28:42:20.6 &     21.4 &     40.8 &       57 &   0.29 &   0.32 &  N \\ 
 7 &  17:46:08.3 -28:41:44.1 &     18.6 &    125.6 &      252 &   0.33 &   0.53 & 17:46:08.4 -28:41:45.6 &     22.6 &     98.8 &      155 &   0.34 &   0.45 &  N \\ 
 8 &  17:46:08.6 -28:43:45.6 &     14.1 &     30.4 &       52 &   0.21 &   0.29 & 17:46:08.2 -28:43:50.1 &     18.8 &     78.7 &      101 &   0.23 &   0.50 &  Y \\ 
 9 &  17:46:08.6 -28:42:09.1 &     31.6 &    111.1 &      216 &   0.55 &   0.38 & 17:46:08.7 -28:42:10.1 &     30.3 &    110.5 &      169 &   0.44 &   0.39 &  Y \\ 
10 &  17:46:09.1 -28:43:18.6 &     14.9 &     53.2 &       96 &   0.24 &   0.45 & 17:46:09.3 -28:43:17.1 &     16.0 &     66.4 &       94 &   0.22 &   0.49 &  Y \\ 
11 &  17:46:09.5 -28:42:07.1 &     26.5 &     61.8 &      122 &   0.47 &   0.29 & 17:46:09.5 -28:42:06.6 &     33.9 &     74.9 &      115 &   0.50 &   0.31 &  N \\ 
12 &  17:46:09.7 -28:43:32.1 &     13.2 &     13.4 &       25 &   0.22 &   0.19 & 17:46:09.6 -28:43:30.6 &     15.5 &     33.0 &       47 &   0.21 &   0.30 &  Y \\ 
13 &  17:46:10.1 -28:42:44.6 &     21.0 &     77.0 &      151 &   0.37 &   0.37 & 17:46:10.1 -28:42:46.1 &     22.1 &     60.2 &       92 &   0.32 &   0.32 &  Y \\ 
14 &  17:46:10.3 -28:43:40.6 &     14.8 &     39.0 &       70 &   0.24 &   0.32 & 17:46:10.4 -28:43:39.6 &     15.1 &     21.5 &       30 &   0.20 &   0.24 &  N \\ 
15 &  17:46:10.3 -28:42:28.1 &     28.4 &    109.9 &      219 &   0.51 &   0.39 & 17:46:10.3 -28:42:28.1 &     25.4 &     93.1 &      146 &   0.38 &   0.39 &  Y \\ 
16 &  17:46:10.6 -28:42:36.1 &     25.0 &     33.9 &       67 &   0.44 &   0.23 & 17:46:10.6 -28:42:36.1 &     22.3 &     30.9 &       48 &   0.33 &   0.23 &  Y \\ 
17 &  17:46:10.6 -28:43:08.1 &     12.6 &     78.8 &      150 &   0.21 &   0.54 & 17:46:10.9 -28:42:58.6 &     16.0 &     72.2 &      109 &   0.23 &   0.46 &  Y \\ 
18 &  17:46:10.6 -28:42:17.6 &     51.7 &     78.4 &      156 &   0.92 &   0.31 & 17:46:10.6 -28:42:17.6 &     58.7 &    118.7 &      185 &   0.87 &   0.39 &  Y \\ 
19 &  17:46:11.2 -28:43:33.1 &     30.8 &    102.5 &      181 &   0.48 &   0.40 & 17:46:10.7 -28:43:31.6 &     15.9 &     42.0 &       59 &   0.22 &   0.32 &  Y \\ 
20 &  17:46:11.2 -28:43:13.6 &     17.2 &     61.8 &      114 &   0.28 &   0.41 & 17:46:11.2 -28:43:09.1 &     14.7 &     63.2 &       92 &   0.20 &   0.42 &  Y \\ 
21 &  17:46:08.0 -28:43:13.6 &     20.8 &     30.3 &       48 &   0.29 &   0.29 &                        &          &          &          &        &        &  Y \\ 
22\tablefootmark{a} &  17:46:08.1 -28:43:50.6 &     18.3 &     38.9 &       63 &   0.27 &   0.27 &                        &          &          &          &        &        &  Y \\ 
23 &  17:46:08.4 -28:41:33.6 &     16.8 &     45.6 &       92 &   0.30 &   0.32 &                        &          &          &          &        &        &  N \\ 
24 &  17:46:09.3 -28:42:26.1 &     19.8 &     25.9 &       50 &   0.34 &   0.22 &                        &          &          &          &        &        &  Y \\ 
25\tablefootmark{a} &  17:46:10.7 -28:43:29.6 &     14.3 &     14.3 &       26 &   0.23 &      U &                        &          &          &          &        &        &  Y \\ 
26\tablefootmark{a} &  17:46:11.2 -28:42:19.6 &     27.4 &     43.0 &       84 &   0.48 &   0.26 &                        &          &          &          &        &        &  Y \\ 
27 &  17:46:11.4 -28:42:14.1 &     20.6 &     28.8 &       55 &   0.35 &   0.22 &                        &          &          &          &        &        &  N \\ 
28 &  17:46:12.0 -28:42:56.6 &     18.8 &     30.4 &       54 &   0.30 &   0.30 &                        &          &          &          &        &        &  Y \\ 
29\tablefootmark{a} &                         &          &          &          &        &        & 17:46:07.4 -28:41:34.6 &     18.7 &     31.2 &       50 &   0.29 &   0.26 &  N \\ 
30 &                         &          &          &          &        &        & 17:46:07.7 -28:41:09.1 &     28.7 &    107.2 &      159 &   0.41 &   0.44 &  N \\ 
31 &                         &          &          &          &        &        & 17:46:08.6 -28:41:16.6 &     22.9 &     22.9 &       34 &   0.33 &      U &  Y \\ 
32\tablefootmark{a} &                         &          &          &          &        &        & 17:46:09.1 -28:41:49.6 &     13.0 &     19.5 &       30 &   0.19 &   0.23 &  N \\ 
33 &                         &          &          &          &        &        & 17:46:09.6 -28:44:01.6 &     26.4 &     86.4 &      113 &   0.33 &   0.44 &  N \\ 
34 &                         &          &          &          &        &        & 17:46:10.4 -28:41:56.1 &     18.2 &     52.7 &       77 &   0.25 &   0.39 &  N \\ 
35\tablefootmark{a} &                         &          &          &          &        &        & 17:46:11.4 -28:42:19.6 &     23.9 &     23.9 &       36 &   0.34 &      U &  N \\ 
36 &                         &          &          &          &        &        & 17:46:11.9 -28:43:31.6 &     28.8 &     33.5 &       44 &   0.36 &   0.23 &  Y \\ 
\label{dendro_contin_table}
\end{tabular}}
\tablefoot{
\tablefoottext{a}{Core 22 is associated with cores 5 and 8. Core 25 is associated with core 19. Core 26 is associated with core 18. Core 27 is partially associated with core 35. Core 29 is associated with core 2. Core 32 is associated with core 7. Core 35 is partially associated with core 27.}
\tablefoottext{b}{Geometric mean diameter of the core. U denotes that the core is unresolved.}
\tablefoottext{c}{Y denotes that the core is associated with 25\,GHz emission above 0.125\,mJy\,beam$^{-1}$ in a 2.2$''$ $\times$ 1.9$''$ beam (Mills et al., in preparation).} 
}
\end{table*}

\subsection{Combined SMA and SCUBA Single Dish Emission \label{SMA_SCUBA_result}}

\begin{figure}
\begin{center}
\includegraphics[width=8cm]{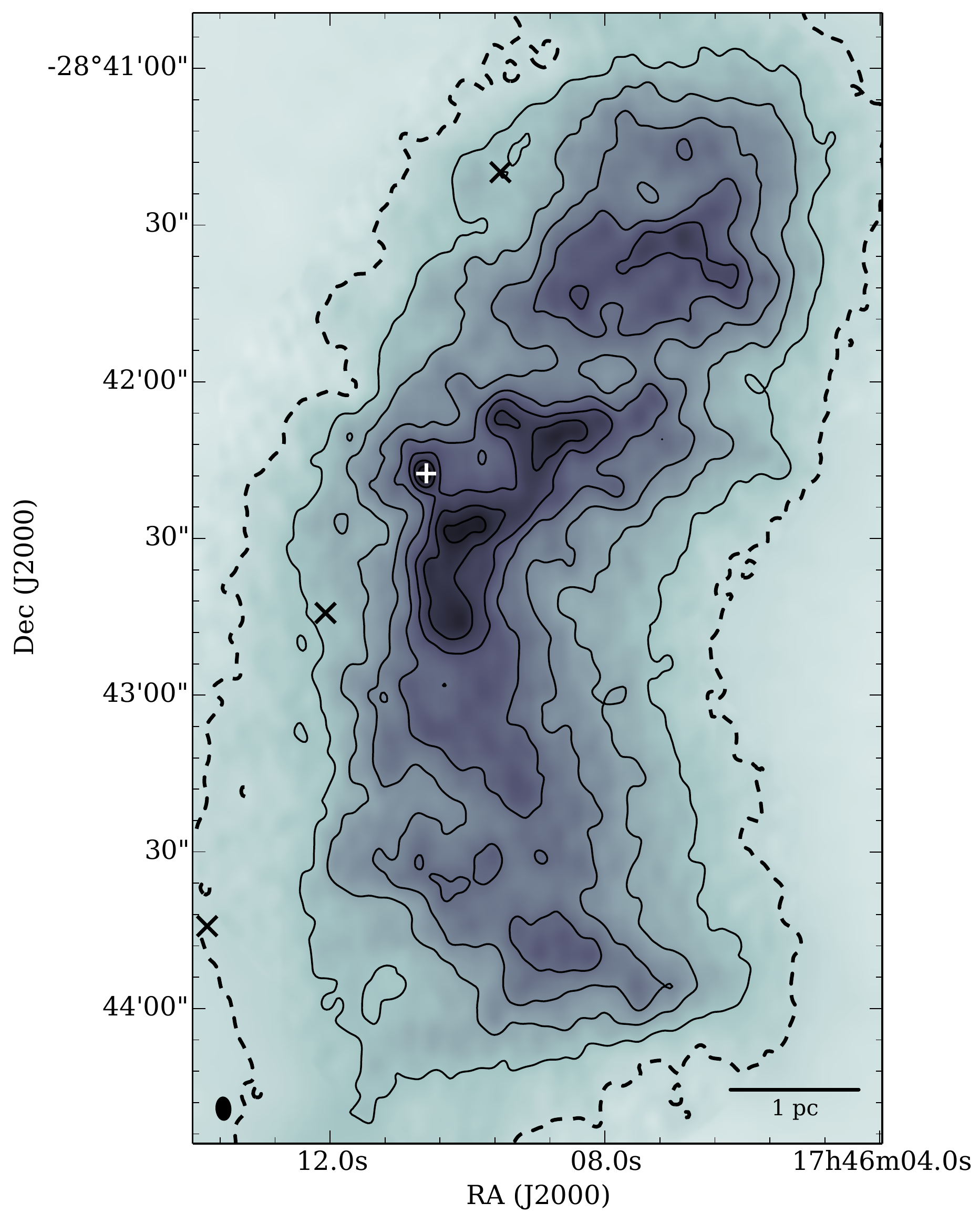}
\caption{Map of the 230.9\,GHz or 1.3\,mm continuum emission observed with the SMA, combined with the scaled single-dish SCUBA 1.3\,mm emission derived in Section~\ref{SD}. Contours are -6, 6, 10, 14, 18, 22, 26 and 30 $\times$ 4\,mJy\,beam$^{-1}$. Greyscale: -20 to 150 mJy\,beam$^{-1}$. The synthesised beam is the same as the SMA-only image, and is shown in the bottom left corner: 4.3$''$ $\times$ 2.7$''$, P.A. = 4.0$^{\circ}$. The plus sign marks the position of the water maser reported by \citet{lis94} and the crosses mark (respectively from north to south) the positions of the 1.3\,cm sources VLA 4 to 6 from \citet{rodriguez13}. The dashed black contour shows a column density of 2$\times10^{22}$\,cm$^{-2}$ for $\beta=$1.75, used as the outer boundary of the cloud when determining its total mass.}
 \label{combined_1.3mm}
 \end{center}
\end{figure}

Figure \ref{combined_1.3mm} presents the combined SMA and scaled SCUBA 1.3\,mm continuum emission derived as described in Section~\ref{SD}. This is one of the few continuum maps that successfully combines the single-dish with interferometer data and hence recovers all spatial scales from the interferometer resolution limit to the scales covered by the bolometer. With this data we can now analyse the column density structure of this enigmatic cloud in great depth. Figure \ref{combined_1.3mm} shows that the three 1.3\,cm sources discovered in \citet{rodriguez13} again appear to be coincident with the edge of the cloud and not any dense, and thus possibly star-forming, material within it. The brightest continuum emission lies between the declinations of -28$^{\circ}$42$'$ and -28$^{\circ}$43$'$. Above a declination of -28$^{\circ}$42$'$, there is also a somewhat separated island of 1.3\,mm emission. There is comparatively less emission toward the south of the cloud.

To obtain an estimate of the cloud mass, the mass corresponding to each pixel in the combined 1.3\,mm image was determined via the same method as described in Section~\ref{SMA_cont}, taking into account the beam size. The total mass was then found by summing these values above the dashed black contour in Figure \ref{combined_1.3mm}, corresponding to a column density of 2$\times10^{22}$\,cm$^{-2}$ for $\beta=$1.75. The mass values determined for the three values of $\beta$ were 10, 9.1 and 7.9 $\times10^{4}$\,M$_{\odot}$ for $\beta=$1.5, 1.75 and 2.0 respectively. Although sightly smaller, this range of values compares well with previous estimates of the cloud mass \citep[M=1.3 - 7 $\times10^{5}$\,M$_{\odot}$,][]{lis94,longmore12, immer12}. 

Summing the masses of Cores 1-20 for the lower or upper side band, as well as Cores 21-36, we determined an estimate of the total mass in cores to be 3388 and 3070\,M$_{\odot}$ respectively, giving a range for the fraction of mass of the cloud in cores between 3.1 and 4.3\%. Unfortunately, it is not possible to compare this directly to the dense gas mass fractions of 10 and 20\% derived respectively by \citet{lada12} and \citet{kainulainen13}, as the magnitude limits that were used by these authors to determine both the total and dense gas mass ($A_{K} >$  0.1 and 0.8\,mag respectively) correspond\footnote{To determine the column densities from $A_{K}$, we assumed $A_{K}$ = 1 mag is equivalent to $A_{\rm v}$ = 8.8 mag \citep{kainulainen13} and $N(H_2) / A_{\rm v} = 0.94 \times10^{21}$\,cm$^{-2}$ mag$^{-1}$ \citep{bohlin78}.} to 8.3$\times10^{20}$\,cm$^{-2}$ and 6.6$\times10^{21}$\,cm$^{-2}$, which are both lower than the lower H$_2$ column density limit used to determine the mass of G0.253+0.016: 2$\times10^{22}$\,cm$^{-2}$.

\subsection{Column Density Probability Distribution Function (PDF) of G0.253+0.016 \label{PDFs}}

The PDF of the volume or column densities within molecular clouds has previously been used as a tool to investigate the effect of various competing physical processes within them \citep[e.g.][]{kainulainen09,schneider13,federrath13}. Often a range of densities within the PDF can be fit well by a log-normal distribution. This distribution of densities is thought to arise from a series of multiplicative, randomly distributed shocks in a turbulent medium, which result in a log-normal volume density distribution due to the central limit theorem \citep{vazquez-semadeni94, ballesteros-paredes11}. When the detected cloud volume densities along the line of sight remain correlated, the column density should represent its mean along the line of sight, and thus will be correspondingly narrower but exhibit the same functional shape. The log-normal form of the column density PDF can be expressed as:

\begin{equation}
p(\eta)~d \eta = p_0(\eta) \exp{ \left[ - (\eta - \mu)^2 / (2 \sigma_{\eta}^2)\right]}~d \eta
\; ,
\label{NPDFeqn}
\end{equation}

where $\eta=ln{(N/\langle N \rangle)}$ with $N$ being the column density; $p_0(\eta)$ is the normalisation constant, which is $p_0(\eta) = 1/ \sqrt{2 \pi} \sigma_{\eta}$ in the case of a purely log-normal distribution; $\mu$ is the mean value of $\eta$ and $\sigma_{\eta}$ is the dispersion. 

In addition to a log-normal distribution of densities, a decreasing power-law density distribution or tail is sometimes observed at higher densities \citep[e.g.][]{kainulainen09,schneider13}. Simulations of turbulent star-forming clouds indicate that this change is primarily due to whether gravitational collapse is occurring in the cloud \citep[][]{klessen00}, however \citet{kainulainen11} also argue that the tail can develop even earlier if the cloud is confined by pressure. Thus investigating the form of the PDF for G0.253+0.016 should provide insight into whether gravitational collapse or star formation is occurring, or if turbulence is the main process shaping the cloud.

\begin{figure}
\begin{center}
\includegraphics[width=9cm]{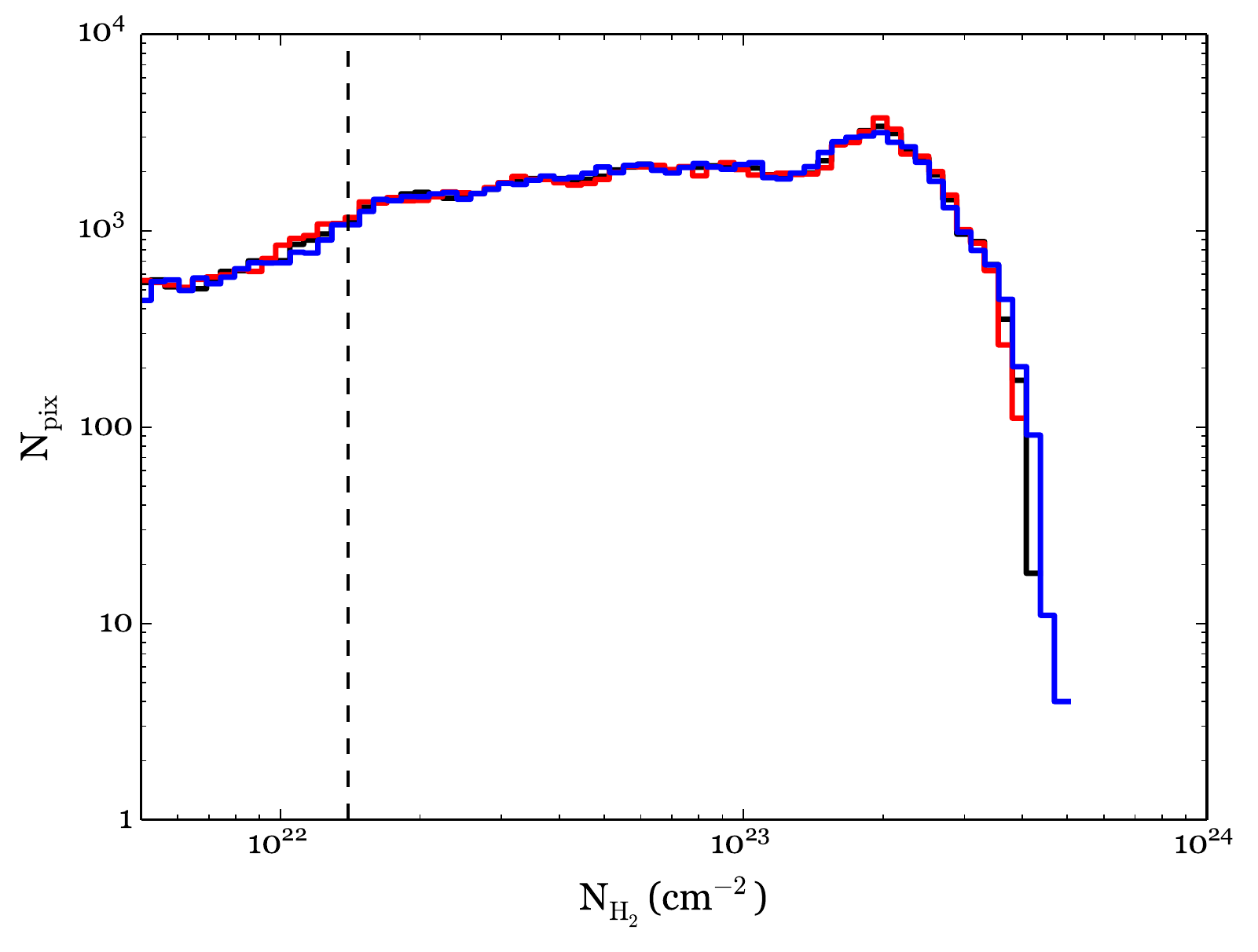}
\caption{Probability distribution functions (PDFs) of column density for G0.253+0.016, assuming three values of $\beta$ when scaling the SCUBA 450$\mu$m emission: 1.5, 1.75 and 2.0 (red, black and blue respectively). The density-values in the PDF shown for $\beta = 1.5$ and 2.0 have been multiplied by factors of 0.87 and 1/0.87 respectively. The PDFs are derived from images with a resolution of 4.3$''$ $\times$ 2.7$''$, or 0.18 $\times$ 0.11\,pc at a distance of 8.4\,kpc. The y-axis shows the number of pixels $N_{\rm pix}$ per logarithmic bin. The dashed line shows the column density corresponding to the 3$\sigma$ noise level for the $\beta=1.75$ image, assuming a temperature for the outer regions of the cloud of 30\,K.}
 \label{PDF_comparebeta}
 \end{center}
\end{figure}

Using the image of the combined 1.3\,mm emission shown in  Figure \ref{combined_1.3mm}, we determined the PDF of column densities in G0.253+0.016. The column density map was determined via the same method as described for the peak column densities in Section~\ref{SMA_cont}, for each pixel in the image. Figure~\ref{PDF_comparebeta} shows the PDF of the column densities from the combined images assuming $\beta$ values of 1.5, 1.75 and 2.0 (red, black and blue respectively). The density-values in the PDF shown for $\beta = 1.5$ and 2.0 have been multiplied by factors of 0.87 and 1/0.87 to show that the overall shape of the PDF is conserved when assuming different values of $\beta$ in scaling the SCUBA 450$\mu$m emission, although there is a small variation between the PDFs seen at the highest densities. The dashed line shows the column density corresponding to the 3$\sigma$ noise level for the $\beta=1.75$ image, assuming a temperature for the outer regions of the cloud of 30\,K. Due to the fact the resultant shapes of PDFs using these three assumptions are very similar for different values of $\beta$ (modulo the shift and thus error of $\pm^{15}_{13}\%$ in absolute value of the column density), we will continue our study using the images and PDF made under the assumption $\beta=1.75$.

\begin{figure}
\begin{center}
\includegraphics[width=9cm]{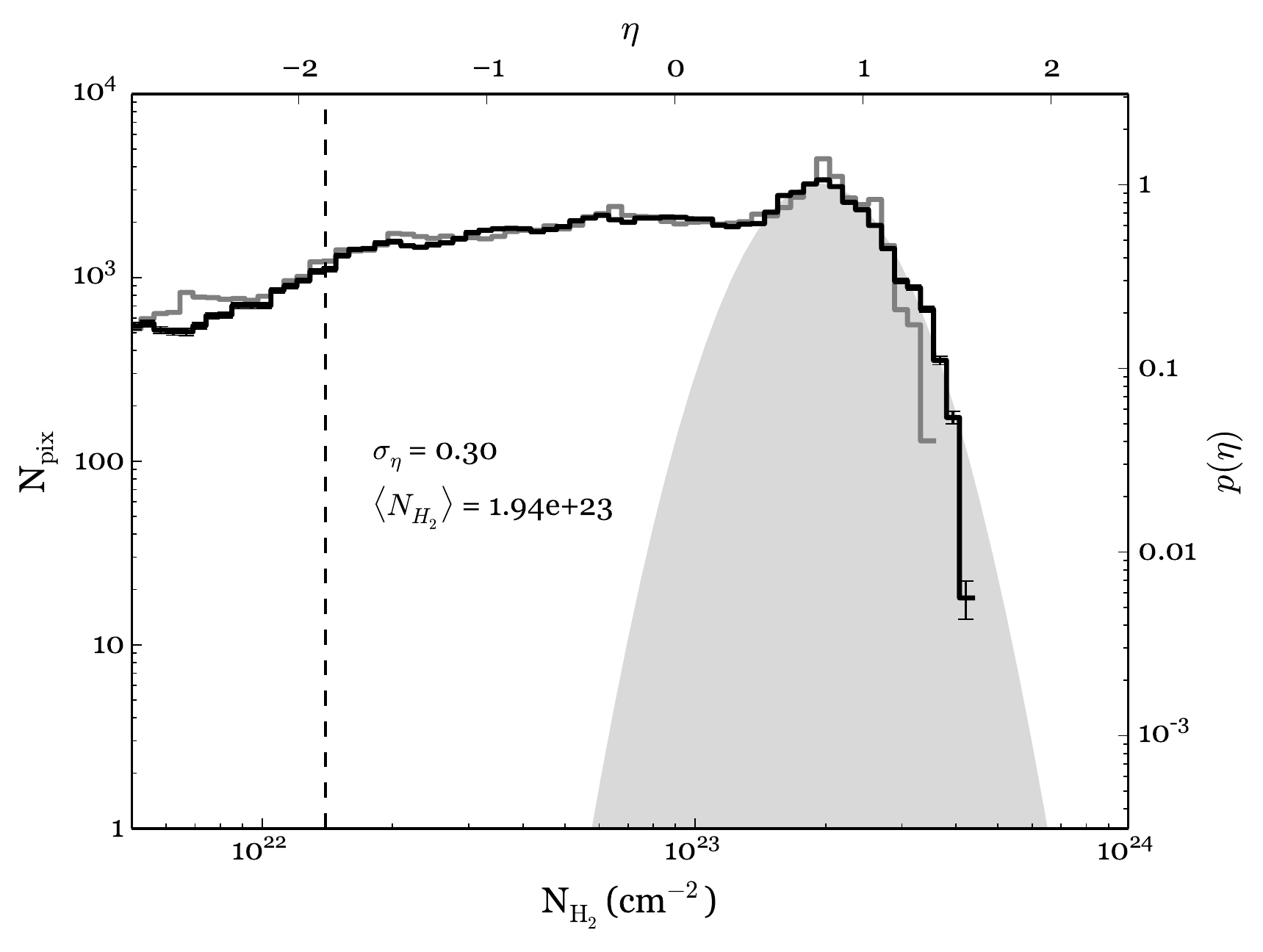}
\caption{Probability distribution functions (PDFs) of the column density for G0.253+0.016, assuming a value of $\beta=1.75$. The black line shows the PDF for the combined image (4.3$''$ $\times$ 2.7$''$ resolution, or 0.18 $\times$ 0.11\,pc for d=8.4\,kpc), and the grey line the SCUBA-only PDF (8$''$ or 0.33\,pc resolution). The left y-axis shows the number of pixels per logarithmic bin $N_{\rm pix}$, and the right y-axis is the normalised probability $p(\eta)$. The top x-axis is displays the dimensionless parameter $\eta=ln{(N/\langle N \rangle)}$. The dashed line shows the column density corresponding to the 3$\sigma$ noise level, assuming a temperature of 30\,K. The Poisson errors are shown as error bars. The grey filled area corresponds to a by-eye fit to the combined PDF, giving the fit parameters $\sigma_{\eta}$, describing the PDF width, and $\langle N_{H_2} \rangle$, the average density, whose values are shown on the Figure.}
 \label{PDF}
 \end{center}
\end{figure}

The grey line in Figure~\ref{PDF} shows the column density PDF of G0.253+0.016 derived from the scaled SCUBA image, and the black line shows the PDF derived  from the combined 1.3\,mm image. In addition to showing the column density and number of pixels in each logarithmic bin on the bottom and left x- and y-axes, we also show the equivalent values of $\eta$ and $p(\eta)$ along the upper and right axes. Both PDFs have a similar shape at low densities: a plateau in the PDF with almost constant probability below a column density of $1.4\times10^{23}$\,cm$^{-2}$. Above this density, the PDFs resemble a log-normal function. Compared to the combined PDF, the SCUBA-only PDF peaks more prominently at the mean density, possibly due to beam-averaging. It also falls below the combined PDF at hydrogen column densities of $\gtrsim3\times10^{23}$\,cm$^{-2}$, which is a further effect of resolution. Similarly, the combined PDF appears to be also altered by the spatial resolution above $\sim4\times10^{23}$\,cm$^{-2}$, where it suffers a sudden drop. This is supported by the fact that the densest continuum core in G0.253+0.016, Core 18, is unresolved, with a column density of 4.4$\times10^{23}$\,cm$^{-2}$. Higher resolution observations would thus ``fill in'' the PDF above this density.

Therefore, up to densities of $4\times10^{23}$\,cm$^{-2}$, our observations show no evidence of a power-law tail in the column density PDF of G0.253+0.016, and therefore we find no indication that the process of collapse and consequent star formation is widespread within the cloud. Nevertheless, we can compare the properties of the log-normal portion of the combined PDF to those found for other clouds, and to predictions from theory, which we discuss in Section \ref{sfpotential}.

We also note that due to the distance of G0.253+0.016, intervening clouds may change its observed PDF, for instance by acting to narrow its true width \citep{schneider14}. However, as we utilised SCUBA observations to determine the lower densities in our PDF, which remove diffuse emission more extended than 4$'$, emission from diffuse foreground and background dust should not strongly affect the observed PDF.

\subsection{SMA Line Emission \label{SMAline}}

\begin{figure*}[!htbp]
\begin{center}
\sidecaption
\includegraphics[width=14cm]{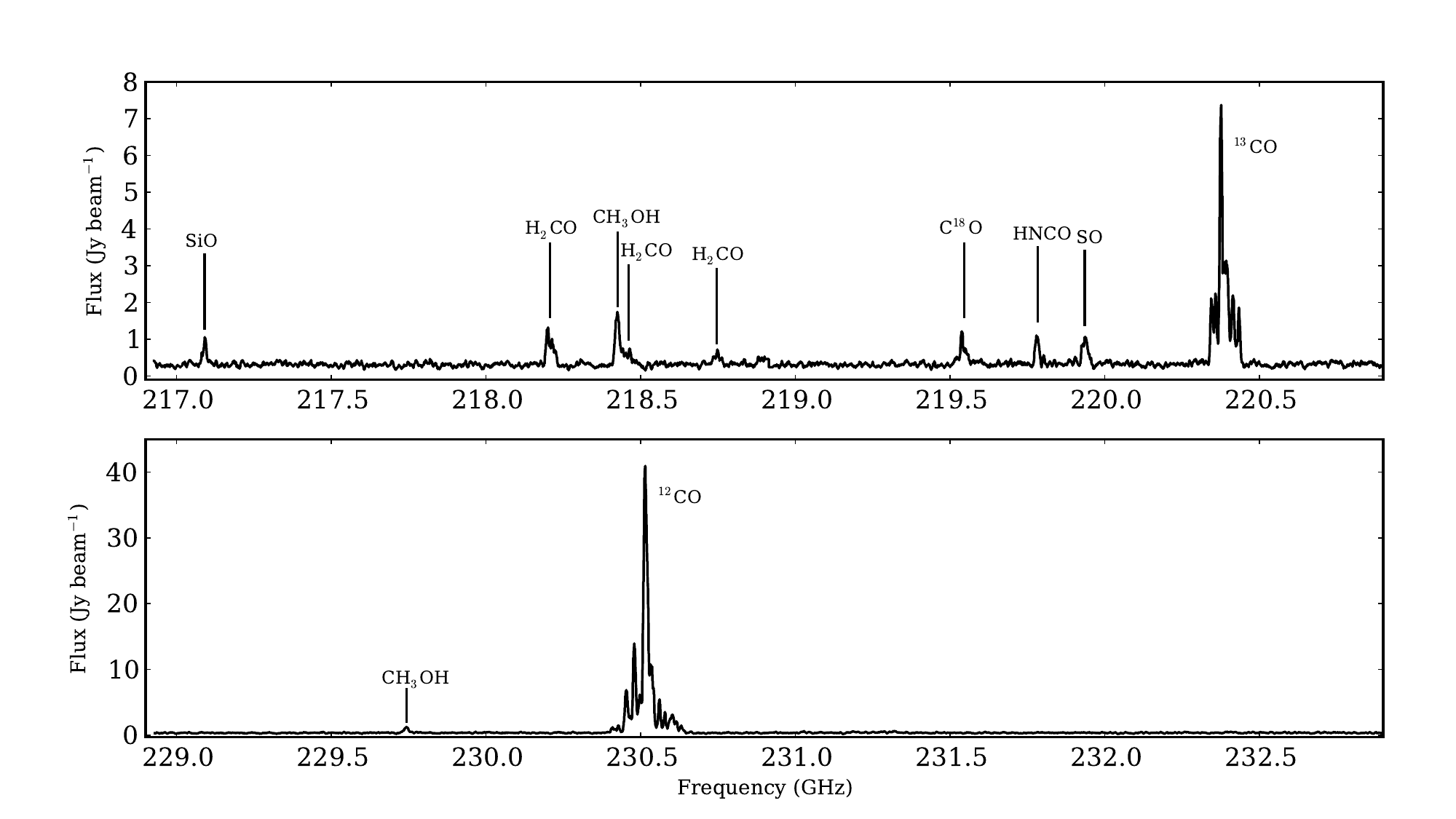}
\caption[]{Observed SMA spectra of both sidebands, averaged over the six observed fields for a short baseline (antennas~1~and~7).}
\label{sideband}
\end{center}
\end{figure*}

Figure \ref{sideband} presents the observed SMA spectra for both sidebands on a short baseline (antennas~1~and~7), averaged over the six observed mosaic fields. The detected lines are marked and are listed in Table \ref{obstable}. Although their frequencies lie within the observed bands, both the dense and hot gas tracers $^{13}$CS(5-4) and the k ladder of CH$_3$CN(12-11) were not detected. The detected lines fall into three groups: shock tracers: methanol, SiO, HNCO and SO; CO isotopologues as diffuse gas tracers: $^{12}$CO, $^{13}$CO and C$^{18}$O; and H$_2$CO lines as temperature probes.

\begin{figure*}
\begin{center}
\includegraphics[width=18.5cm]{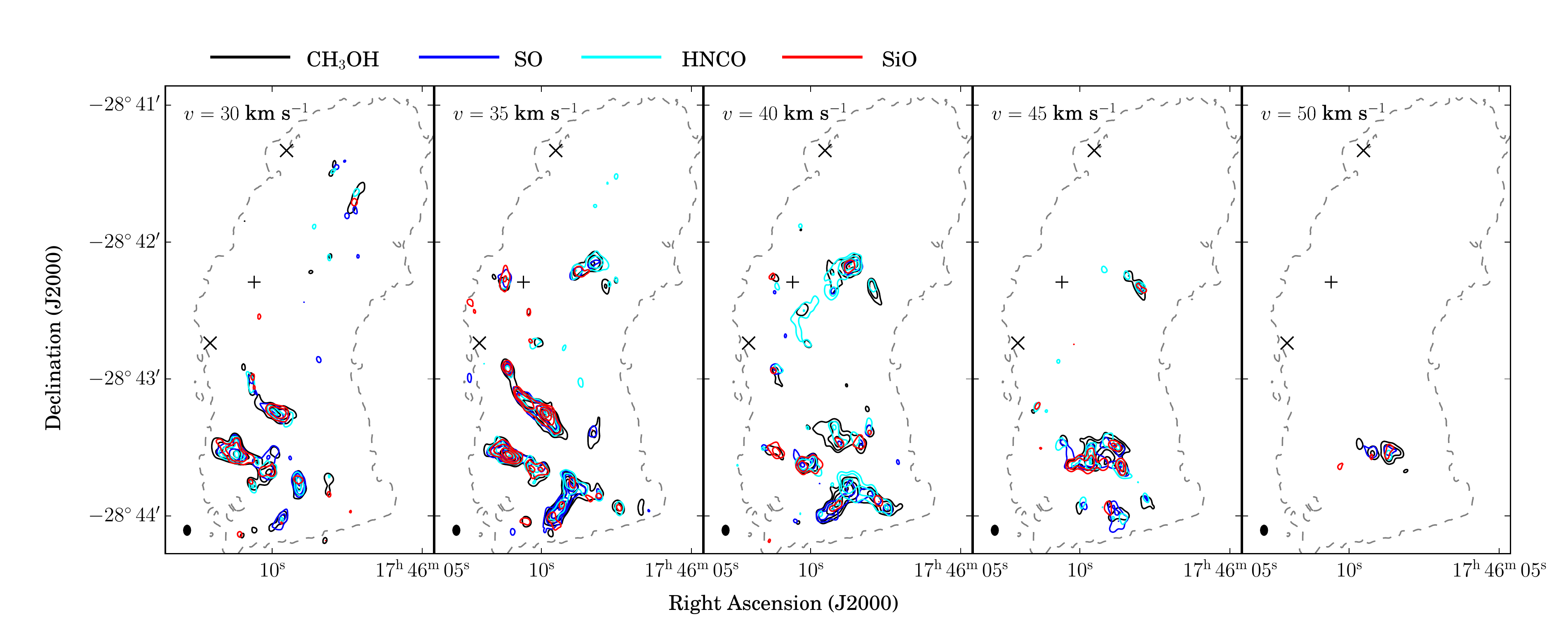}
\caption{SMA CH$_{3}$OH 4(2,2) - 3(1,2)-E, SO, HNCO and SiO emission. Contours are -5, 5, 8, 12, 16 and 20 $\times$ the rms noise values for each line given in Table \ref{obstable}, for a spectral resolution of 5\,kms$^{-1}$. A synthesised beam of 4.3$''$ $\times$ 2.9$''$, P.A. = 0$^{\circ}$ is shown in the bottom left corner. The plus sign marks the position of the water maser reported by \citet{lis94} and the crosses mark (respectively from north to south) the positions of the 1.3\,cm sources VLA 4 and 5 from \citet{rodriguez13}. The dashed grey contour shows the combined dust continuum emission at a level of 0.024\,mJy\,beam$^{-1}$, the lowest solid black contour shown in Figure~\ref{combined_1.3mm}.}
 \label{SMA_CH3OH_SiO_SO}
 \end{center}
\end{figure*}

Figure~\ref{SMA_CH3OH_SiO_SO} presents the channel maps of the shock tracing lines. This includes the brightest observed methanol line, CH$_{3}$OH 4(2,2) - 3(1,2)-E at 218.44\,GHz (the other methanol lines observed also present emission similar to that seen in Fig.~\ref{SMA_CH3OH_SiO_SO}), as well as SiO, HNCO and SO. In all transitions, the morphology is similar with the emission being brightest towards the southern half of G0.253+0.016. This is also seen for the H$_2$CO and the CO isotopologue emission. In addition, a similar morphology is seen in collisionally excited methanol masers at 36\,GHz \citep[][Mills et al. in preparation]{mills13b}. In Figure \ref{SMA_CH3OH_SiO_SO}, three spoke-like filaments can be seen, which are most prominent in the 30 and 35\,km\,s$^{-1}$ channels, radiating out from an apex lying at approximately 17$^{\rm h}$46$^{\rm m}$08$^{\rm s}$ -28$^{\circ}$43$'$36$''$ (J2000). In the 45\,km\,s$^{-1}$ channel, the dominant emission moves westward toward the apex and becomes ring-like. In the 50\,km\,s$^{-1}$ channel it becomes more compact and continues to move toward the apex point. This behaviour suggests a large velocity gradient (spanning at least 20\,km\,s$^{-1}$) along these filaments. Figure \ref{ch3oh_mom1} presents the first moment map of the 4(2,2) - 3(1,2)-E methanol line confirming this, where we have marked the three filaments with dashed lines. In section \ref{dynbrick} we discuss the possible mechanisms to produce such a velocity gradient. In Figure \ref{ch3oh_mom1}, the 1.3\,mm SMA dust continuum emission is also shown in black contours. Although they do not directly line-up with one another, there is an apparent spatial correlation between the methanol and dust continuum emission. 

Figure~\ref{compare_line_cont} compares the 1.3\,mm dust continuum cores and line emission integrated from 0 to 60\,km\,s$^{-1}$ for several lines. The line emission is that observed with the SMA, apart from $^{13}$CO which also incorporates the IRAM 30m emission, which will be described further in the following section. The brightest 1.3\,mm source (Core 18) is not obviously detected in any of the integrated images, although it lies just above 5\,$\sigma$ in C$^{18}$O in the 40\,km\,s$^{-1}$ channel. Otherwise, there is line emission consistently offset to the east of core 18 (e.g. in H$_2$CO where the emission is coincident with  Core 26). Other examples of coincident emission are CH$_3$OH, SiO, HNCO and H$_2$CO associated with Cores 4, 6 and 24, and CH$_3$OH, SiO and HNCO with Core 9 (the CO isotopologues are also coincident with Core 9 but do not peak there). The CO isotopologue maps have a bar of strong emission in the south at a declination of approximately -28$^{\circ}$43.5$^{'}$, which is strongly correlated with Cores 12, 14, 19, 25 and 36. In general there is a good agreement between the morphology of the line and dust continuum emission, although not all dust cores have a distinct counterpart in the line emission.

\subsection{Combined SMA and IRAM 30m Emission \label{combinedSMA_IRAM}}

\begin{figure}
\includegraphics[width=9cm]{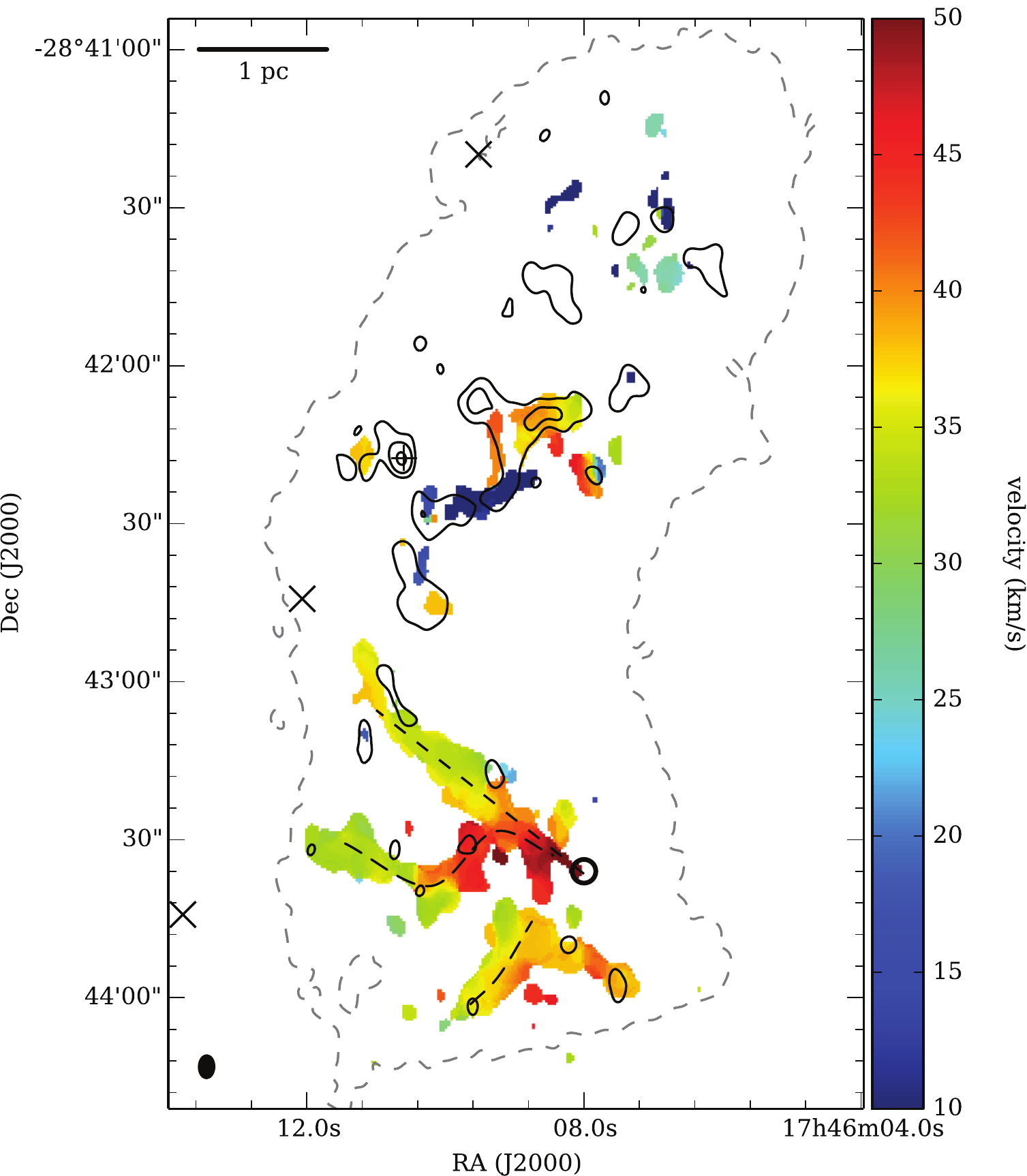}
\caption{First moment map of the 4(2,2) - 3(1,2)-E methanol line at 218.440\,GHz. Black contours show the 230.9\,GHz or 1.30\,mm continuum emission observed with the SMA at 5, 10 and 20 $\times$ rms noise = 2.5\,mJy\,beam$^{-1}$. The synthesised beam for the methanol line is shown in the bottom left corner: 4.3$''$ $\times$ 2.9$''$, P.A. = -1.1$^{\circ}$. The plus sign marks the position of the water maser reported by \citet{lis94} and the crosses mark (respectively from north to south) the positions of the 1.3\,cm sources VLA 4 to 6 from \citet{rodriguez13}. The thick black circle in the south of the image marks the position of point A and the dashed lines show the positions of the three filaments mentioned in the text. The dashed grey contour shows the combined dust continuum emission at a level of 0.024\,mJy\,beam$^{-1}$, the lowest black contour shown in Figure~\ref{combined_1.3mm}. \label{ch3oh_mom1}}
\end{figure}

\begin{figure*}
\begin{center}
\includegraphics[width=15cm]{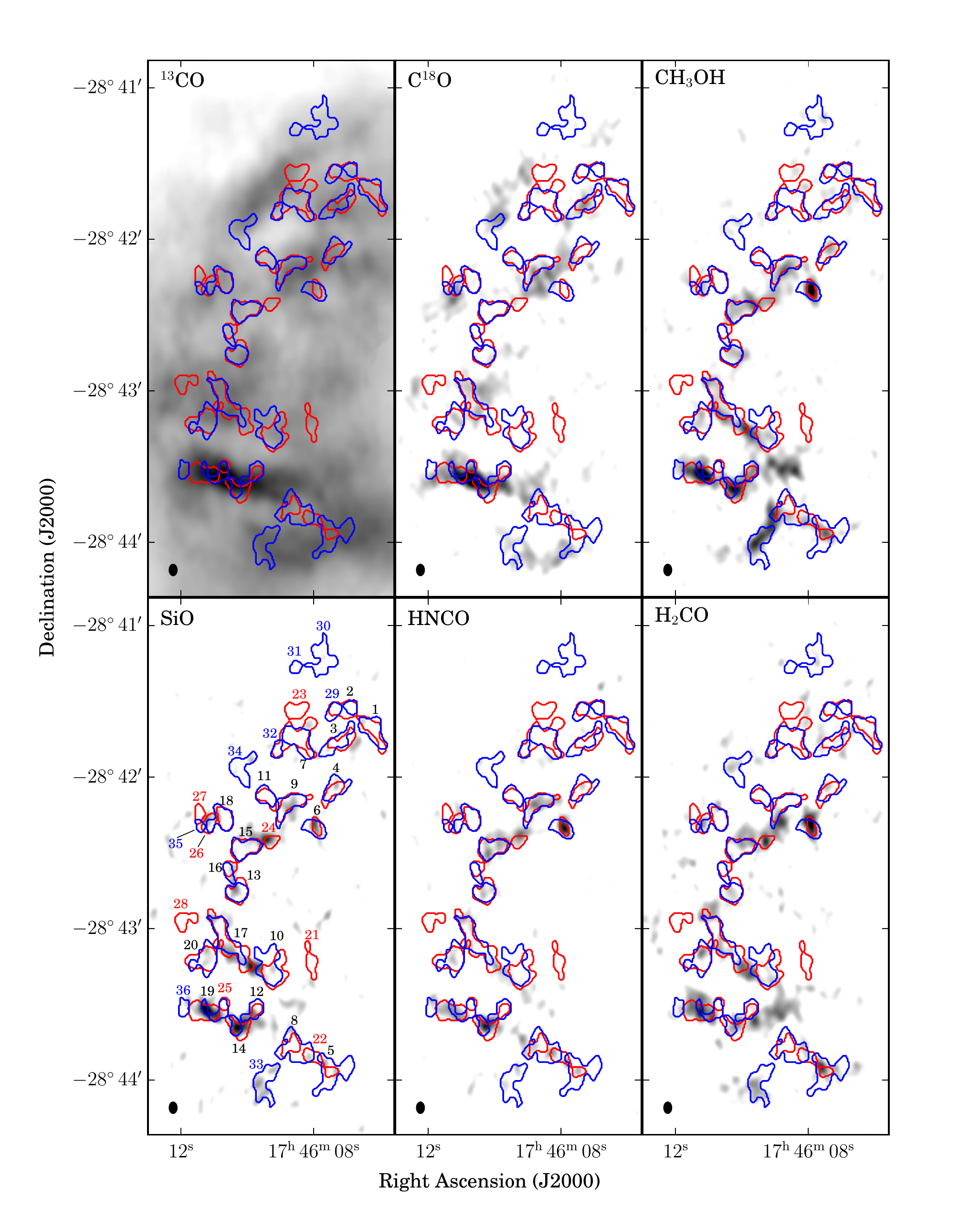}
\caption{Comparison between dust continuum and line emission integrated between 0 and 60\,kms$^{-1}$. The red and blue contours show the positions of the cores detected in SMA dust continuum (blue and red for 1.3 and 1.37\,mm respectively, see Fig.~\ref{SMAcontinuum_numbered}). A representative synthesised beam of 4.3$''$ $\times$ 2.9$''$, P.A. = 0$^{\circ}$ is shown in the bottom left corner. The greyscale shows the SMA line emission (combined with IRAM 30m in the case of $^{13}$CO) integrated between 0 and 60\,kms$^{-1}$ for six lines: $^{13}$CO, C$^{18}$O, CH$_{3}$OH 4(2,2) - 3(1,2)-E, SiO, HNCO and H$_{2}$CO 3(0,3) - 2(0,2). The lines are shown with stretches 5-10\,mJy\,beam$^{-1}$\,kms$^{-1}$ for $^{13}$CO, 3-14\,mJy\,beam$^{-1}$\,kms$^{-1}$ for C$^{18}$O and CH$_3$OH, and 3-10\,mJy\,beam$^{-1}$\,kms$^{-1}$ for the remaining lines.}
 \label{compare_line_cont}
 \end{center}
\end{figure*}

\begin{figure*}
\begin{center}
\includegraphics[width=19cm]{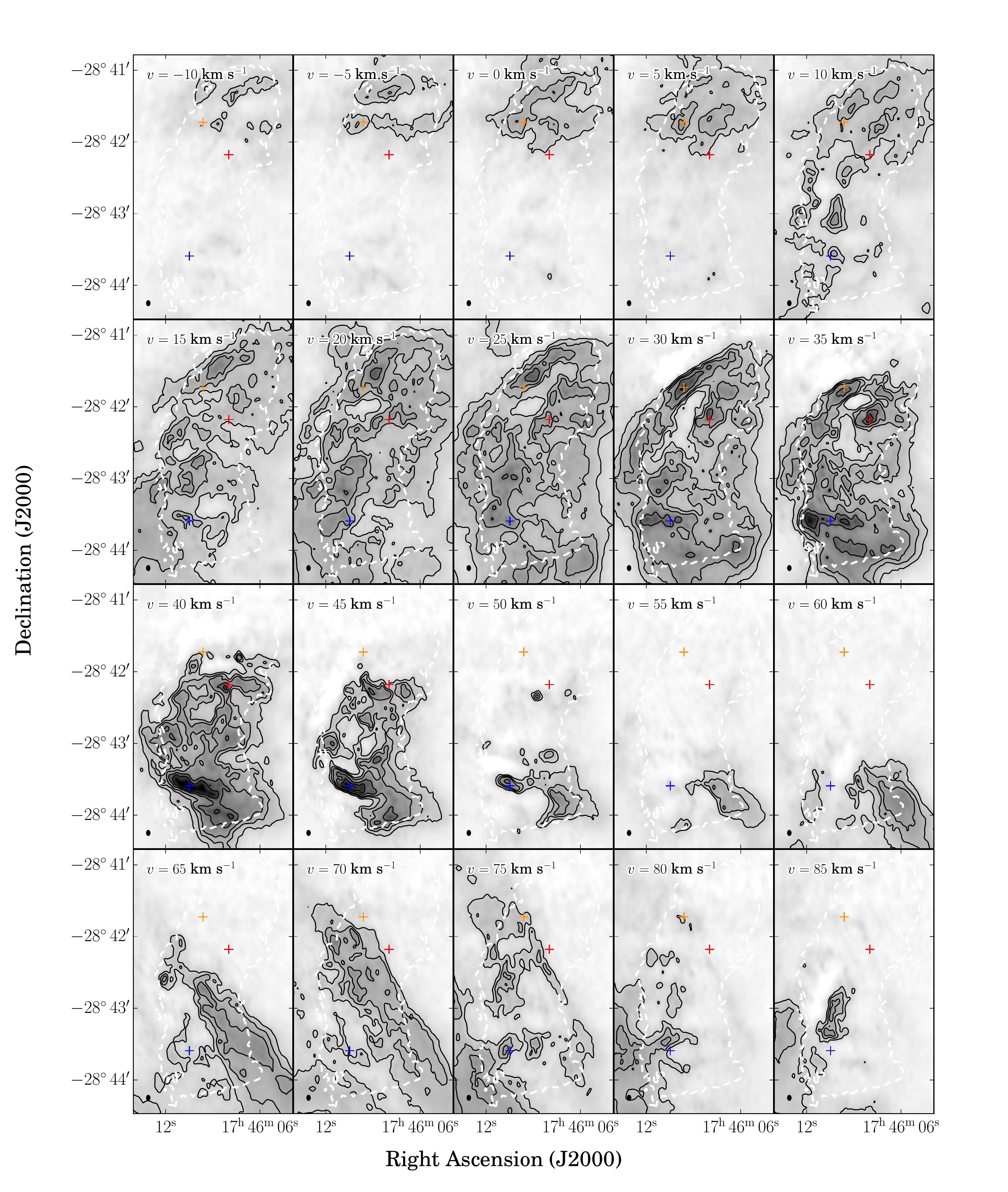}
\caption{Combined SMA + IRAM 30m $^{13}$CO emission between -10 and 85\,kms$^{-1}$. Contours are 10, 15, 20, 30 and 40 $\times$ the map sensitivity 0.14\,mJy\,beam$^{-1}$, for a spectral resolution of 5\,kms$^{-1}$. The greyscale ranges from -0.14 to 8\,mJy\,beam$^{-1}$. The beam size of 4.2$''$ $\times$ 2.9$''$, P.A. = 0.7$^{\circ}$ is shown in the bottom left corner, which corresponds to the size of the SMA beam. The dashed white contour shows the combined 1.3\,mm dust continuum emission at a level of 0.024\,mJy\,beam$^{-1}$, the lowest black contour shown in Fig.~\ref{combined_1.3mm}. The coloured crosses correspond to the positions of the spectra shown in Fig.~\ref{13CO_spectra}. \label{IRAM13COfig}}
\end{center}
\end{figure*}

Figure \ref{IRAM13COfig} presents the combined SMA and IRAM 30m $^{13}$CO emission observed towards G0.253+0.016, imaged in 5\,km\,s$^{-1}$ channels. The emission from G0.253+0.016 extends between approximately -10 to 55\,km\,s$^{-1}$. In addition, there is another cloud which can be seen between 60 and 85\,km\,s$^{-1}$, which covers the south-east half of the map. We will refer to this as the 70\,km\,s$^{-1}$ cloud. There is a large velocity gradient across G0.253+0.016, from blueshifted in the north to redshifted in the south, which has been previously noted by \citet{lis98} and \citet{rathborne14}. Between 15 and 35\,km\,s$^{-1}$, and most obvious in the 30 and 35\,km\,s$^{-1}$ channels, there is a cavity in the $^{13}$CO emission in the north half of the cloud, which is seen at approximately 17$^{\rm h}$46$^{\rm m}$09$^{\rm s}$.5 -28$^{\circ}$42$'$30$''$ (J2000). This hole or cavity may be due to optical depth effects, as the combined 1.3mm continuum shows the brightest emission and thus densest material close to this position. The morphology of G0.253+0.016 also appears curved, with its sections of emission connecting to form a semi-circular bow directed to the east. The emission can also be seen to curve in a similar manner in the SMA continuum and line emission in Fig.~\ref{compare_line_cont}. Between 40 and 50\,km\,s$^{-1}$, the bar of emission in the south of the cloud has a velocity gradient which increases to the west, similarly to the gradient seen in the lines presented in Figs. \ref{SMA_CH3OH_SiO_SO} and \ref{ch3oh_mom1}. Between 50 and 85\,km\,s$^{-1}$, the morphology of the $^{13}$CO emission connects smoothly to the 70\,km\,s$^{-1}$ cloud, where it first contracts to a small area of emission in the south of the cloud in the 55\,km\,s$^{-1}$ channel, and then expands from the same point into the diagonal bar of emission which can be seen in the 65 and 70\,km\,s$^{-1}$ channels.

\begin{figure}
{\center
\includegraphics[width=9.5cm]{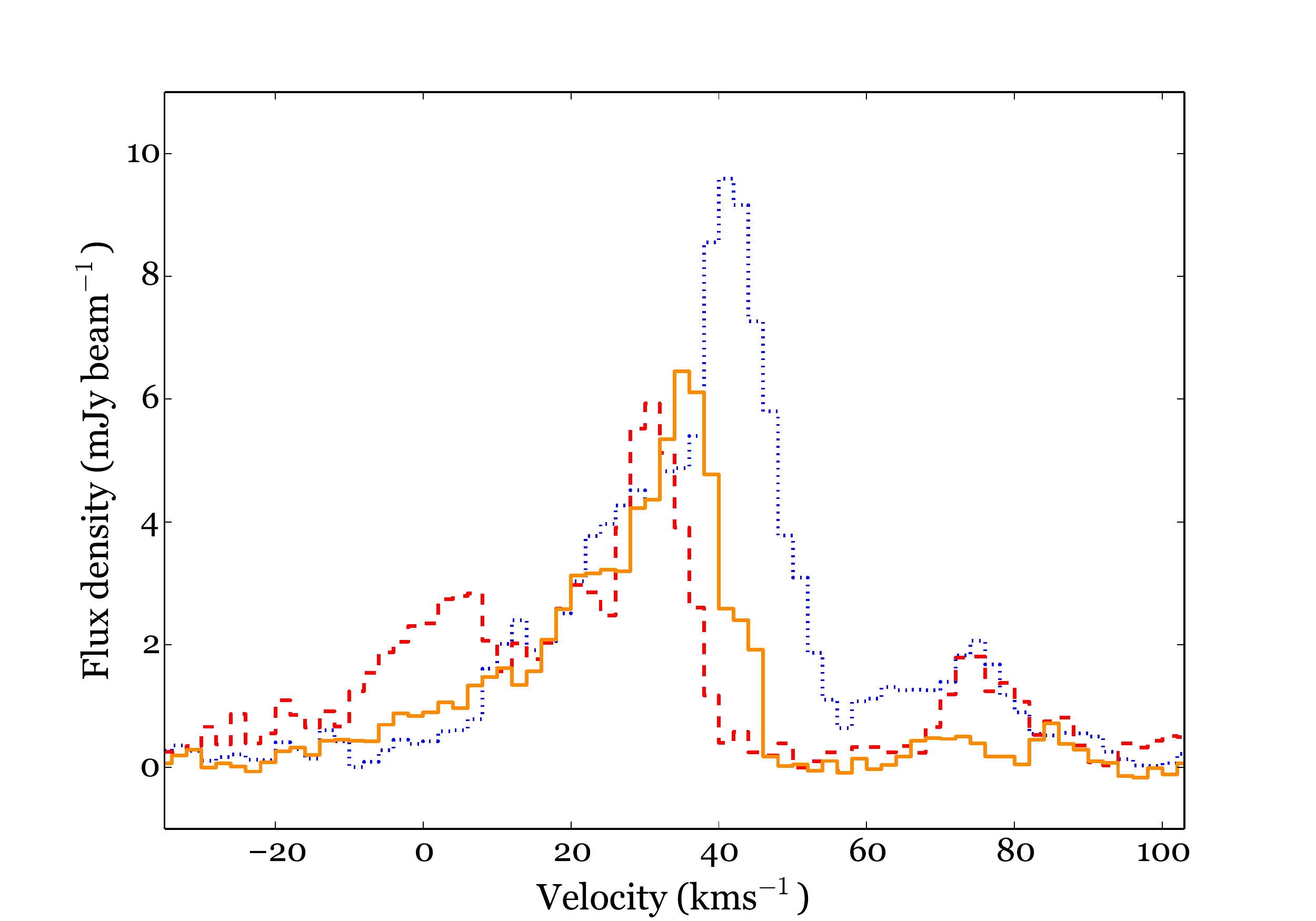}
\caption{Spectra of the combined SMA plus IRAM 30\,mm $^{13}$CO emission at three positions in the map: 17$^{\rm h}$46$^{\rm m}$10.5$^{\rm s}$ -28$^{\circ}$43$'$35$''.5$ (blue dotted), 17$^{\rm h}$46$^{\rm m}$08.0$^{\rm s}$ -28$^{\circ}$42$'$10$''.7$ (red dashed), and 17$^{\rm h}$46$^{\rm m}$09.6$^{\rm s}$ -28$^{\circ}$41$'$43$''.5$ (orange solid, all J2000). The spectral resolution is 2\,km\,s$^{-1}$. The three positions corresponding to these spectra are marked with crosses of the same colour in Figure \ref{IRAM13COfig}.\label{13CO_spectra}}}
\end{figure}

Figure \ref{13CO_spectra} displays combined SMA plus IRAM 30\,mm $^{13}$CO spectra at three selected positions in the cloud, which are marked with crosses of the same colour in Fig. \ref{IRAM13COfig}. The positions are (ordered by increasing declination): 17$^{\rm h}$46$^{\rm m}$10.5$^{\rm s}$ -28$^{\circ}$43$'$35$''.5$ (blue dotted line), 17$^{\rm h}$46$^{\rm m}$08.0$^{\rm s}$ -28$^{\circ}$42$'$10$''.7$ (red dashed line), and 17$^{\rm h}$46$^{\rm m}$09.6$^{\rm s}$ -28$^{\circ}$41$'$43$''.5$ (orange solid line, all J2000) Although the three spectra are broad and offset in velocity, they all display a relatively narrow peak at respectively 41.8, 30.6 and 34.6\,km\,s$^{-1}$. Fitting the brightest components of the spectra with a gaussian line profile, we found that the linewidths of this narrower component to be 13.9, 11.4 and 14.1\,km\,s$^{-1}$. In addition to the bright (relatively) narrow components, there is also non-gaussian emission which extends to lower velocities. Emission from the 70\,km\,s$^{-1}$ cloud can be seen to varying degrees in each spectrum.

Figure \ref{13COPV} presents a position velocity (PV) diagram collapsed along the Right Ascension axis of the cube for $^{13}$CO in greyscale and grey contours, overplotted with CH$_3$OH tracing shocks in yellow contours. We chose CH$_3$OH as it was the brightest shock-tracing molecule, however SiO also shows a similar morphology in PV space. Both G0.253+0.016 and the 70\,kms$^{-1}$ cloud can be clearly seen in $^{13}$CO emission. The methanol emission is only seen towards G0.253+0.016 and not the cloud at 70\,kms$^{-1}$. As previously noted, the shock-tracing emission is concentrated towards the south of G0.253+0.016, and spans the largest velocity range in this region: from $\sim$25 to 55\,kms$^{-1}$ at a declination of -28$^{\circ}$43$'$33$''$ (J2000). The shock emission at this declination also reaches out along a bridge of $^{13}$CO emission to the  70\,kms$^{-1}$ cloud. The peak of the $^{13}$CO emission lies between 30 and 45\,kms$^{-1}$, behind the velocity of the arrowhead of methanol emission pointing into the other cloud. We discuss the possibility that these features are evidence of a collision between G0.253+0.016 and the 70\,kms$^{-1}$ cloud in Section \ref{dynbrick}. At higher declinations, there are two velocity components: one at $\sim$10 and one at $\sim$40\,kms$^{-1}$. By comparison with Figs.~\ref{ch3oh_mom1} and \ref{compare_line_cont}, we can see that the 10\,kms$^{-1}$ component is associated with Cores 15 and 24, and the 40\,kms$^{-1}$ component with Core 9. Whilst these two methanol velocity components with associated dust cores are very close on the sky, they are separated by 30\,kms$^{-1}$. Thus although they lie within the confines of the larger G0.253+0.016 cloud traced by $^{13}$CO, there is evidence for strong dynamics in the north of the cloud. It is also interesting to note that the $^{13}$CO emission peaks at a velocity of $\sim$25\,kms$^{-1}$, lying in between these two methanol velocity components.

\subsection{Kinetic temperatures from H$_2$CO \label{h2coSect}}

The ratio of the integrated fluxes of H$_2$CO lines can be used to determine the kinetic temperature of the gas \citep[e.g.][]{mangum93}. For example, recent single-dish observations of the CMZ using H$_2$CO line ratios have determined average temperatures of $\sim$65\,K \citep{ao13}. These gas temperatures are significantly higher than the measured dust temperatures in the CMZ \citep[e.g. 21$\pm$2 \,K,][]{pierce-price00} and in G0.253+0.016 ($<$30\,K from Section \ref{SD}). The cause of these different dust and gas temperatures may be explained by an increased level of heating by cosmic rays and the interstellar radiation field in the Galactic centre \citep[e.g.,][]{clark13}, or instead by shocks \citep[e.g.,][]{martin-pintado97}. Here we investigate the temperatures traced by our SMA H$_2$CO observations.

As the line H$_2$CO 3(2,2) - 2(2,1) was blended with a methanol line, we instead used the H$_2$CO 3(0,3) - 2(0,2) and 3(2,1) - 2(2,0) transitions, hereafter H$_2$CO line 1 and 2, which have lower state energies of 10.5 and 57.6\,K, and upper state energies of 21.0 and 68.1\,K respectively \citep[Cologne Database for Molecular Spectroscopy,][]{muller01}. To obtain a ratio map of line 1 over line 2, we summed the flux in both H$_2$CO 5\,kms$^{-1}$ resolution image cubes which was above 1$\sigma$ in each image. We then took the ratio of the two lines, using only the summed flux values above 4$\sigma$: 0.6\,Jy\,beam$^{-1}$\,kms$^{-1}$. Figure \ref{h2coratio} shows the ratio map of line 1 over line 2. For all positions where it was possible to calculate the line ratio, the value is less than 3, with the average line ratio being approximately 1.4. 

To investigate the temperatures and number densities which would be required to produce these ratios, we used a Python wrapper for RADEX \citep{van-der-tak07} written by A. Ginsburg\footnote{https://code.google.com/p/agpy/wiki/Radex}, assuming a uniform sphere geometry. We determined an estimate for the peak column density of H$_2$CO by multiplying the peak dust column density of Core 19 reported in Table~\ref{dendro_contin_table}, 0.48\,g\,cm$^{-2}$ or N$_{H_2}$ = 1.0$\times10^{23}$\,cm$^{-2}$, by an H$_2$CO abundance for the Galactic centre relative to H$_2$ of 1.2$\times$10$^{-9}$ \citep{ao13}. This provided a value of N$_{H_2}$ = 1.2$\times10^{14}$\,cm$^{-2}$. However, this value may be lower, as the H$_2$CO abundance for G0.253+0.016 reported by \citet{guesten83} was only 3$\times$10$^{-11}$, a value within a factor of two of the abundance in Galactic disk clouds, which gives n(H$_2$) = 3$\times10^{12}$\,cm$^{-2}$. We did not use dust cores with higher peak column densities as they were not detected in H$_2$CO. We also used the column densities derived from the SMA-only data given in Table~\ref{dendro_contin_table}, not the combined SMA plus SCUBA continuum data, as the measured SMA column densities correspond to similar spatial scales as those probed by the SMA line data.

By fitting the spectrum associated with each pixel with gaussians, the FWHM linewidths of both H$_2$CO lines were determined to range between 2 and 19\,kms$^{-1}$. Figure \ref{h2co_temp} shows the dependence of the line ratio on kinetic temperature and volume density for an assumed H$_2$CO column density of $10^{14}$\,cm$^{-2}$ and a linewidth of 10\,kms$^{-1}$. The yellow contour shows a value of 1.4, indicating that temperatures above 370\,K are able to produce the observed line ratios. For a linewidth of 2\,kms$^{-1}$ the derived temperatures are $>$500\,K, and for a linewidth of 19\,kms$^{-1}$, $>$345\,K. At lower column densities of $10^{13}$\,cm$^{-2}$, the temperature and volume density dependencies change little with the assumed linewidth, resulting in temperatures above $\sim$320\,K, while at column densities of $10^{15}$\,cm$^{-2}$, temperatures above 500\,K are required.

For an H$_2$CO column density of $10^{14}$\,cm$^{-2}$, the volume density for a ratio of 1.4 is centred around 2$\times$10$^{4}$\,cm$^{-3}$. Given the peak column density above, N$_{H_2}$ = 1.0$\times10^{23}$\,cm$^{-2}$, this volume density is expected for a line of sight depth of 1.6\,pc or 40$''$ at a distance of 8.4\,kpc. Thus, as the largest angular scale recovered by the data is only $\sim$20$''$, it is likely that the true volume density is slightly higher, and/or the representative column densities are somewhat lower than 1.0$\times10^{23}$\,cm$^{-2}$.

In summary, our results from H$_2$CO line ratios show that the kinetic temperatures in G0.253+0.016 reach upward of $\sim$320\,K, much higher than the temperatures of the dust ($<$30\,K). These gas temperatures are also higher than the estimates from single-dish formaldehyde observations \citep[e.g. $\sim$65\,K,][]{ao13}, however we suggest that this is because the denser molecular gas in G0.253+0.016 is hotter on the smaller scales traced by the SMA ($\sim$0.15\,pc). This is supported by \citet{mills13}, who uncovered a high-temperature component of $\sim$330\,K in G0.253+0.016 using multi-level Green Bank Telescope observations of inversion transitions of ammonia up to NH$_3$(13,13), providing strong constraints on the gas temperature. In addition, \citet{rodriguez-fernandez01} observed 16 CMZ clouds including G0.253+0.016 in H$_2$ pure-rotational lines with the \textit{Infrared Space Observatory} \citep[ISO,][]{kessler96} and discovered a warm ($\sim$150\,K), diffuse (10$^{3}$\,cm$^{-3}$) component in these clouds traced by the H$_2$ S(0) and S(1) transitions, and in five of their observed clouds, a hot ($\sim$600\,K), dense ($\lesssim$10$^{6}$\,cm$^{-3}$) component traced by the H$_2$ S(5) and S(4) transitions. Unfortunately, the H$_2$ S(5) and S(4) transitions were not observed by \citet{rodriguez-fernandez01} for G0.253+0.016 (denoted as M+0.24+0.02 in their work), however they derived a temperature of 163$\pm$2\,K and a column density of 1.73$\pm$0.06$\times10^{22}$\,cm$^{-2}$ for the warm component from the detected S(0) and S(1) transitions. Our higher resolution observations may therefore be tracing the more dense and hot gas associated with the higher H$_2$ transitions seen in the other clouds.

\section{Discussion \label{discussion}}

\subsection{Internal dynamics in G0.253+0.016 \label{dynbrick}}
 
In Section \ref{SMAline} we presented CH$_3$OH line observations which displayed a large velocity gradient (spanning at least 20\,km\,s$^{-1}$) in the Southern part of G0.253+0.016. Possible explanations of such a gradient include accretion along the three observed filaments in this region (marked with dashed lines in Fig.~\ref{ch3oh_mom1}), which accelerate away from the observer toward a gravitational potential, or acceleration of the gas by a cloud-cloud collision. In the first case, the measured velocity gradient along the filaments can be used to determine the central mass required to accelerate the gas, which would most likely lie close to the apex where the filaments converge at highest redshift near 17$^{\rm h}$46$^{\rm m}$08$^{\rm s}$ -28$^{\circ}$43$'$36$''$ (J2000, henceforth point A, marked as a thick circle in Fig.~\ref{ch3oh_mom1}). Figure~\ref{ch3oh_pv} shows a position-velocity cut along the northernmost filament, measured between 17$^{\rm h}$46$^{\rm m}$08$^{\rm s}$.0 -28$^{\circ}$43$'$36$''$.5 and 17$^{\rm h}$46$^{\rm m}$11$^{\rm s}$.0 -28$^{\circ}$43$'$05$''$.4 (J2000) and shown as the longest, straight, dashed line in Fig.~\ref{ch3oh_mom1}. The velocities were fit at radii between 0.25 and 1.5\,pc (assuming a distance of 8.4\,kpc) with a simple function describing the infall velocity, equal to the escape velocity, with an additional shift to account for the average cloud velocity, 

\begin{equation}
{\rm v} = \sqrt{\frac{2 G M}{R}} + {\rm v}_{_{\rm LSR}}
\end{equation}

\noindent The fit is shown as a dashed black line in Fig.~\ref{ch3oh_pv}. The fit shown requires a mass $M = 4\times10^{4}$\,M$_{\odot}$ and an offset velocity v$_{_{\rm LSR}}=19$\,km\,s$^{-1}$. This provides a order-of-magnitude estimate of the mass which would be required to accelerate the gas. Taking a conservatively large estimate of 10$''$ or 0.41\,pc for the radius containing this mass around point A, we find an integrated mass of 1600$\pm^{15}_{13}$\%\,M$_{\odot}$ within this region, which is $<$5\% of that required to accelerate the gas. Hence, this scenario can not explain the observed velocity gradient.

\begin{figure*}
\begin{center}
\includegraphics[width=15cm]{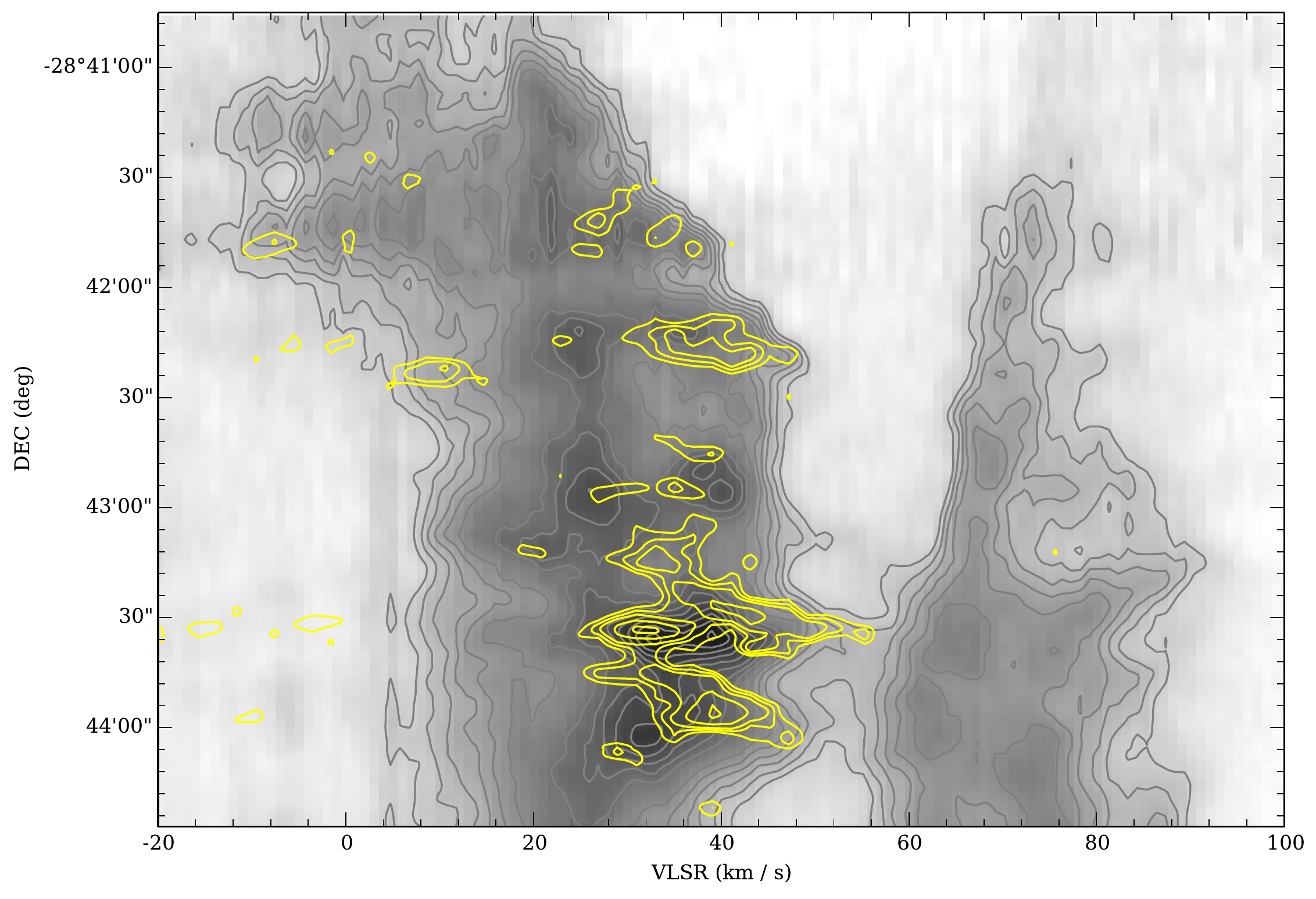}
\caption{Position velocity diagram of Declination against velocity for the combined $^{13}$CO SMA + IRAM 30m in greyscale and contours and CH$_3$OH SMA emission in yellow contours. The emission has been collapsed (summed) along the Right Ascension axis of the cube. \label{13COPV}}
\end{center}
\end{figure*}

\begin{figure}
\includegraphics[width=8.5cm]{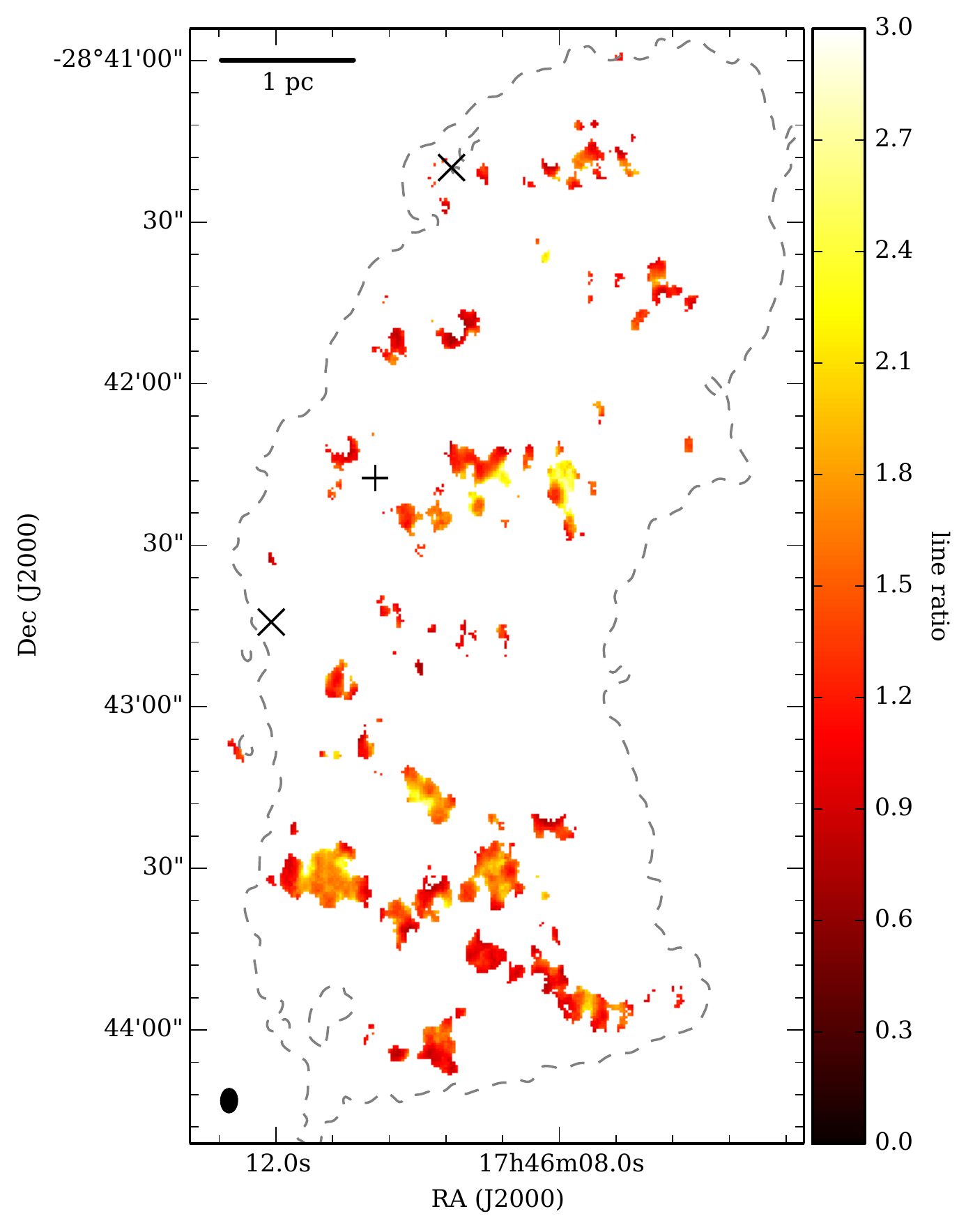}
\caption{The integrated flux line ratio of the 3(0,3) - 2(0,2) over 3(2,1) - 2(2,0) transition of H$_2$CO. The synthesised beam for both lines and the ratio image is shown in the bottom left corner: 4.3$''$ $\times$ 2.9$''$, P.A. = -1.0$^{\circ}$. The plus sign marks the position of the water maser reported by \citet{lis94} and the crosses mark (respectively from north to south) the positions of the 1.3\,cm sources VLA 4 and 5 from \citet{rodriguez13}. The dashed grey contour shows the combined dust continuum emission at a level of 0.024\,mJy\,beam$^{-1}$, the lowest black contour shown in Figure~\ref{combined_1.3mm}. \label{h2coratio}}
\end{figure}

\begin{figure}
\includegraphics[width=9.cm]{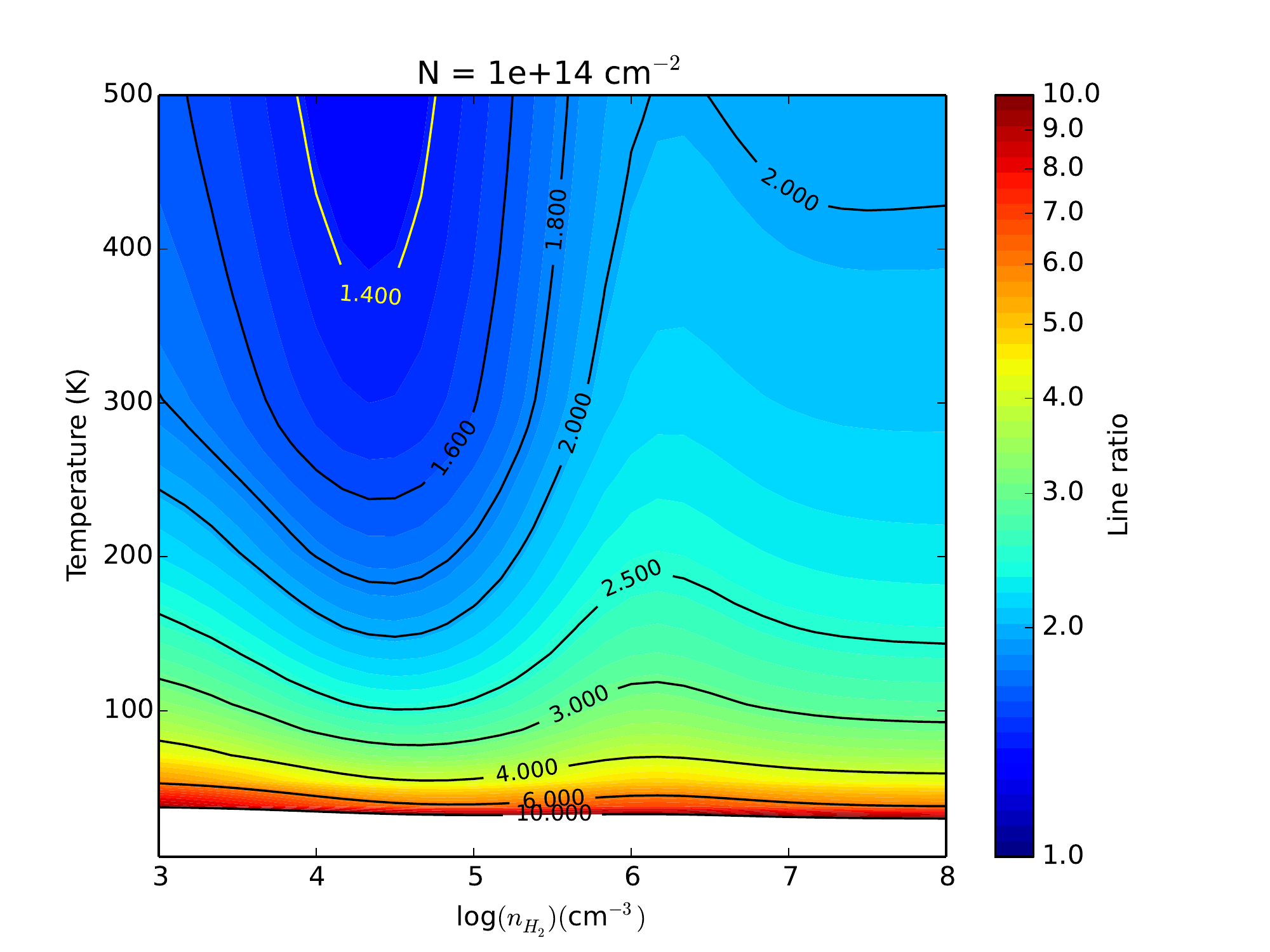}
\caption{Dependency of the H$_2$CO3(0,3) - 2(0,2) over 3(2,1) - 2(2,0) line ratio on kinetic temperature and volume density for an assumed column density of $10^{14}$\,cm$^{-2}$ and a linewidth of 10\,kms$^{-1}$. The black contours show the line ratio at the labelled intervals. The yellow contour shows a line ratio value of 1.4. \label{h2co_temp}}
\end{figure}

\begin{figure}
{\center
\includegraphics[width=7cm]{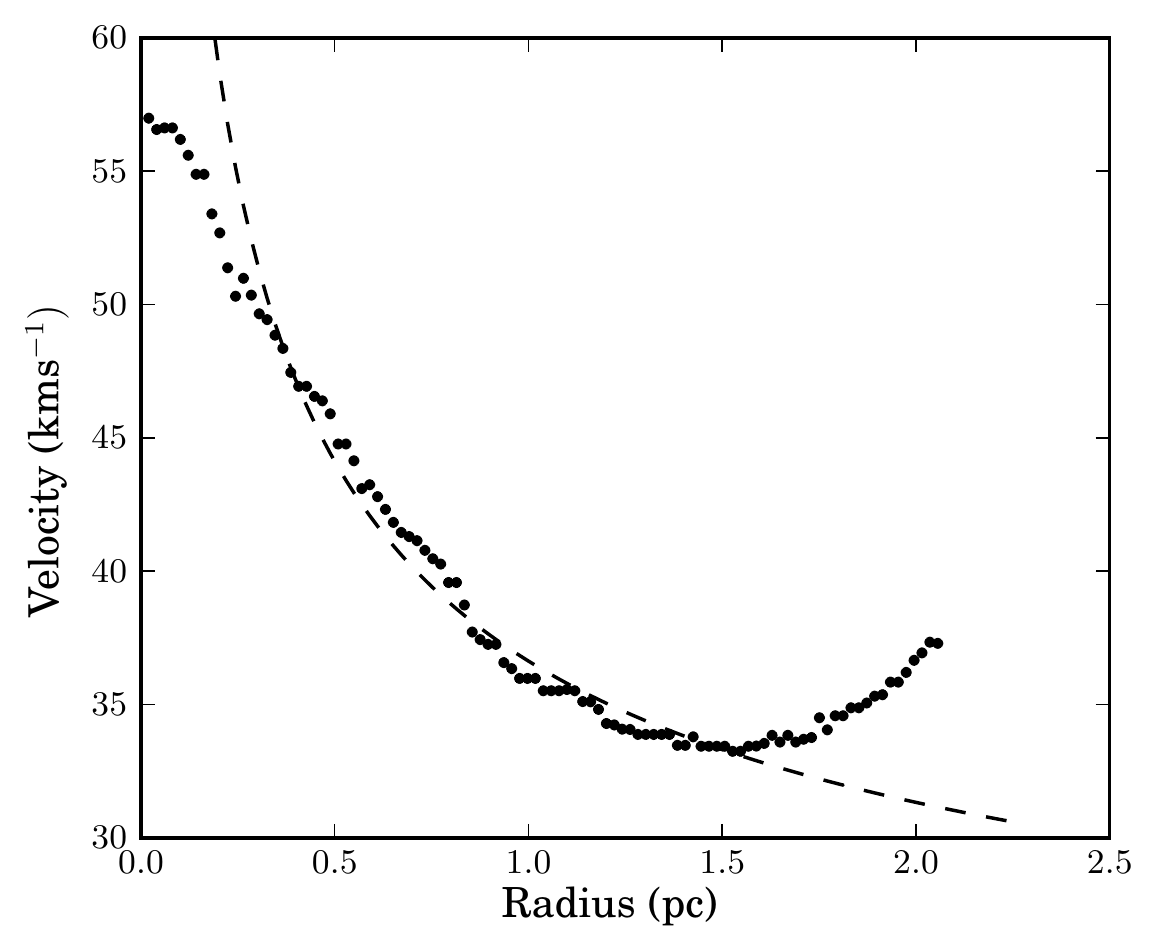}
\caption{Filled circles show a position-velocity cut along the northmost filament shown in Fig.~\ref{ch3oh_mom1}, between 17$^{\rm h}$46$^{\rm m}$08$^{\rm s}$.0 ${-28}^{\circ}$43$'$36$''$.5 and 17$^{\rm h}$46$^{\rm m}$11$^{\rm s}$.0 ${-28}^{\circ}$43$'$05$''$.4 (J2000). The dashed line presents a fit to the data between radii of 0.25 and 1.5\,pc (assuming a distance of 8.4\,kpc).} \label{ch3oh_pv}
}
\end{figure}

An alternative explanation is that a cloud-cloud collision is causing the velocity gradient in the molecules shown in Figure~\ref{SMA_CH3OH_SiO_SO}. Methanol, SO, SiO and HNCO, are commonly seen in shocks \citep[e.g.][]{menten09, pineau-des-forets93, martin-pintado97, rodriguez-fernandez10}.  In fact, as the results of \citet{menten09} and \citet{martin-pintado97} show, widespread emission from shock tracers is a common property of the CMZ. In Section \ref{combinedSMA_IRAM}, we presented a position-velocity diagram of $^{13}$CO and CH$_3$OH (Fig. \ref{13COPV}). The figure shows that the $^{13}$CO emission from G0.253+0.016 and the 70\,km\,s$^{-1}$ cloud overlap in the south of G0.253+0.016, suggesting that these two clouds may be interacting. Figure \ref{13COPV} also shows that the brightest broad-linewidth CH$_3$OH emission is found close to this overlap region. In fact, two strips of broad-linewidth CH$_3$OH emission appear to reach from G0.253+0.016 into the 70\,km\,s$^{-1}$ cloud across their intermediate velocities. Thus we interpret the above signatures as evidence that G0.253+0.016 is colliding with another cloud in the CMZ at 70\,km\,s$^{-1}$. Cloud collision has previously been suggested by \citet{lis01} and \citet{higuchi14} as a scenario to explain the shock tracers seen in G0.253+0.016, and here we identify the likely partner cloud in this collision. Such cloud-cloud collisions may be a method to create dense star clusters \citep[e.g.,][]{fukui14,higuchi14}. For instance, the bow-shock morphology of the larger more massive cloud after collision shown in the cloud-cloud collision simulations of \citet[][see their Figure 1]{anathpindika10} bears resemblance to the curved morphology of the continuum and line emission seen in Figs.~\ref{compare_line_cont} and~\ref{IRAM13COfig}.

\subsection{The interaction of G0.253+0.016 with its environment \label{dynenv}}

\begin{figure*}
\begin{center}
\sidecaption
\includegraphics[width=14cm]{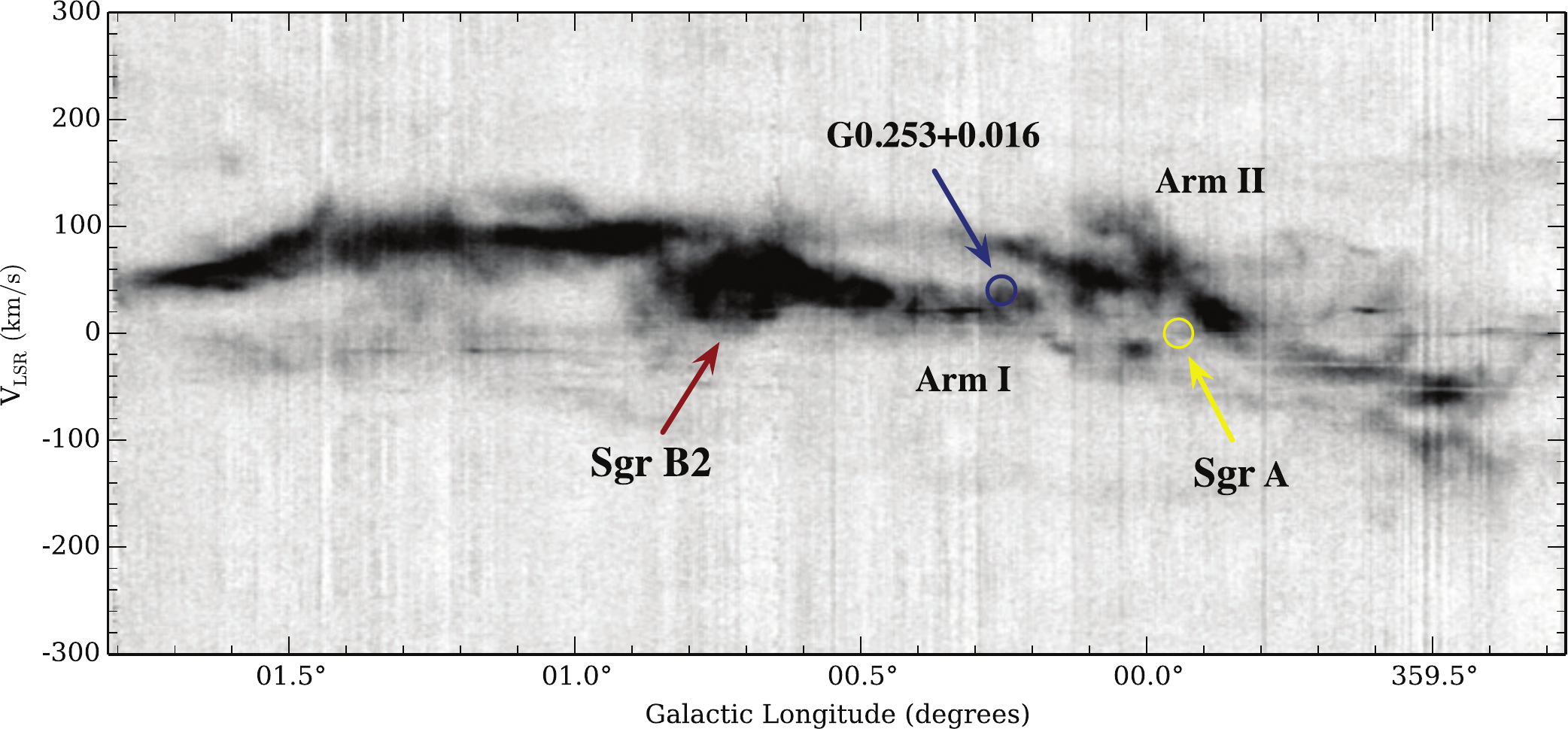}
\caption[]{Galactic longitude-velocity diagram of the CMZ in HNC, using the 3\,mm line data presented in \citet{jones12}. To make the $\ell$-v diagram, the emission was integrated between Galactic latitudes $b$ = -0.29 and 0.21. The approximate positions and velocities of G0.253+0.016, Sgr\,A and Sgr\,B2 are shown, and Arms I and II from \citet{sofue95} are labelled.}
\label{HNC_pv}
\end{center}
\end{figure*}

To investigate the interaction of G0.253+0.016 with its environment, we used the HNC molecular line data of the Central Molecular Zone from the Mopra CMZ molecular line survey, presented in \citet{jones12}. Figure~\ref{HNC_pv} presents a Galactic longitude-velocity diagram similar to their Figure 7. We used the HNC line as it suffered from less absorption or striping in the $\ell$-v diagram. To make the $\ell$-v diagram, the emission was integrated between Galactic latitudes $b$ = -0.29 and 0.21$^{\circ}$. The approximate positions and velocities of G0.253+0.016, Sgr\,A and Sgr\,B2 are shown. Two coherent structures can be seen in Fig.~\ref{HNC_pv}, corresponding to Arms I and II originally named by \citet{sofue95}. Arm I stretches from $\ell$=359.4$^{\circ}$ to Sgr\,B2, and Arm II covers the full length of the observed Galactic longitudes, and is redshifted with respect to Arm I . Although Arm II still can be seen, it becomes faint at longitudes between Sgr\,B2 and G0.253+0.016. If these two Arms formed one coherent structure, then this shape in $\ell-$v space between $\ell$=359.4$^{\circ}$ and Sgr\,B2 would denote an elliptical orbit of gas. As G0.253+0.016, which lies in Arm I, is seen as a dark cloud against the bright background of the CMZ, it is likely that it is positioned on the near side of the CMZ. The fact that Arm I is blueshifted with respect to Arm II indicates that Arm I is also the mostly approaching section of the elliptical orbit while Arm II is the mostly receding one. This result agrees with the orientation of the CMZ determined by \citet{sawada04}, who found, by the comparison of CO emission and OH absorption data, that the CMZ is an elongated structure orientated $\sim$70$^{\circ}$ with respect to the line of sight, with the Galactic eastern end being closer to us. In this picture, the gas should be redshifted after passing Sgr B2, and blueshifted after passing Sgr C at $\ell \sim359.4^{\circ}$, as seen in the observed longitude-velocity diagram. This is also in agreement with dynamical models of the CMZ, which find that, due to the fact the gas experiences pressure and viscous forces, the central $x_2$ orbits of the bar potential describing the CMZ should only lead the bar or $x_1$ orbits by $\sim$45$^{\circ}$ \citep[see Figure 2 of][for a schematic of this orientation, and for a full review of dynamical models of the CMZ]{ferriere07}. Further, trigonometric parallax \citep{reid09} as well as modelling of the 6.4\,keV X-ray emission from the Sgr B clouds as a light echo from Sgr A* \citep{ryu09} find Sgr B2 to be closer to the Sun than Sgr A*. This is opposite to the orientation of the CMZ suggested by \citet{molinari11}, which instead predicts that the gas should be blueshifted soon after passing Sgr B2, and redshifted after passing Sgr C, which is not observed.

Having determined the orientation and dynamics of the CMZ in comparison to previous works, we now examine how G0.253+0.016 may interact with its environment. In Section \ref{dynbrick} we presented evidence that G0.253+0.016 was interacting with a cloud at 70\,km\,s$^{-1}$. Looking at Fig.~\ref{HNC_pv}, we see that this cloud would lie in Arm II. In fact, when inspecting an HNC $\ell$-v-b cube of the data, we see that the emission associated with G0.253+0.016 reaches to higher velocities, while the 70\,km\,s$^{-1}$ cloud in Arm II reaches to lower velocities and touches G0.253+0.016. However, if the CMZ consists of one stable orbit of gas, it would be impossible for these two clouds to interact, as they would be on opposite sides of the elliptical orbit. Therefore we suggest that a possible solution to this contradiction could be that Arms I and II are instead two distinct, coherent velocity streams, such as two spiral arms, which follow on from the dust lanes tracing the intersection of $x_1$ orbits in the bar. Arm II eastward of Sgr\,B2 is  probably tracing the inner section of one of the dust lanes of the Milky Way, which follows on to the rest of Arm II beginning at Sgr\,B2. Examples of galaxies with such inner spiral arms and complex nuclear structures are NGC\,5806\footnote{http://www.spacetelescope.org/images/potw1235a/} and NGC\,1300\footnote{http://hubblesite.org/newscenter/archive/releases/2005/01/image/a/}. Therefore we suggest that the interaction of G0.253+0.016 with its environment may require a different structure for the CMZ than an elliptical ring of gas, possibly several spiral arms which interact in a complex manner. 

\subsection{The current star-formation activity and potential of G0.253+0.016}
\label{sfpotential}
In Section \ref{PDFs} we presented the column density PDF of G0.253+0.016. In this section, we compare the observed PDF to that expected from theory. In the case of magnetised turbulence, the dispersion in the three-dimensional density PDF $\sigma_s^2$ should be described as \citep{padoan11,molina12}:

\begin{equation}
\sigma_s^2 = \ln \left[ 1 + b^2 \mathcal{M}_s^2 ~ \beta/ (\beta +1) \right] ,
\label{theoryPDF}
\end{equation}

\noindent where $b$ is the compressive to total power in the turbulent driving, $\mathcal{M}_s $ is the gas Mach number, and $\beta$ is the ratio of gas to magnetic pressure. We assumed $b=0.4$, corresponding to a `natural' mixture of solenoidal and compressive driving. In the case of 3D turbulence, this is produced by a ratio of compressive to solenoidal turbulence forcing power of 1/2 \citep{federrath10}. We also calculated $\mathcal{M}_s $ for the case of isothermal gas as $\mathcal{M}_s = \sqrt{3} \sigma_{\rm v} / c_s $ where we assumed the one-dimensional velocity dispersion was $\sigma_{\rm v} = 3.1$\,km\,s$^{-1}$, derived from the mean linewidth of H$_2$CO 3(0,3) - 2(0,2) of $\sim$7.4\,km\,s$^{-1}$. The sound speed $c_s$ was determined using

\begin{equation}
c_s = \sqrt{\frac{k_B~T}{2.8~m_H}}\,.
\end{equation}

The temperature was assumed to be 320\,K, and $k_B$ and $m_H$ are the Boltzmann constant and the mass of hydrogen respectively. Thus we found a sound speed of 0.97\,km\,s$^{-1}$ and a Mach number of 5.6. Note, however, that our calculation of the expected magnetic field strength $B$ below is only sightly sensitive to the assumed value of $T$, as for small values of $\beta$, the $c_s^2$ term in the expression for Mach number in Equation \ref{theoryPDF} effectively cancels with that in the expression for $\beta$.

The value of $\beta$ can be determined via $\beta = 8 \pi \rho c_s^2 / B^2$, where $\rho$ is the density of the gas (both ions and neutrals) assuming they are well-coupled, and $B$ is the magnetic field strength measured in gauss. We assumed $\rho = 2.8 m_H n$ where $n=2\times10^4$\,cm$^{-3}$ from our analysis in Section \ref{h2coSect}.

Finally, we assumed the correspondence between the dispersions of the three- and two-dimensional PDFs $\sigma_s = \xi \sigma_{\eta}$, where $\xi=2.7\pm0.5$ \citep{brunt10}.

Given our measured value of $\sigma_{\eta} = 0.30$ (Section \ref{PDFs}) for G0.253+0.016, we expect a magnetic field strength $B = $0.31\,mG is required to produce the observed PDF. This is in agreement with measured values of the magnetic field strength of dense clouds within the inner 200\,pc of the Galaxy, which range from several 0.1\,mG to a few mG \citep{ferriere09}. Further, we propagated our expected uncertainties in $b$ (= 0.33 to 1.0), $\xi$ (= 2.2 to 3.2), linewidth (= 2 to 19\,km\,s$^{-1}$) and $n$ ($\sim 10^4$ to $10^5$\,cm$^{-3}$), finding 3\,$\sigma$ limits for $B$ of 0.031 and 4.6\,mG, again in good agreement with the literature. We note that observations may favour non-compressive turbulence with values of $b$ closer to 1/3 \citep{kainulainen13}, which are covered by our assumed values. Thus our observed column density PDF agrees well with current theory on how the density structure of molecular clouds is determined via turbulence and magnetic fields. 

\begin{figure}
\includegraphics[width=9.cm]{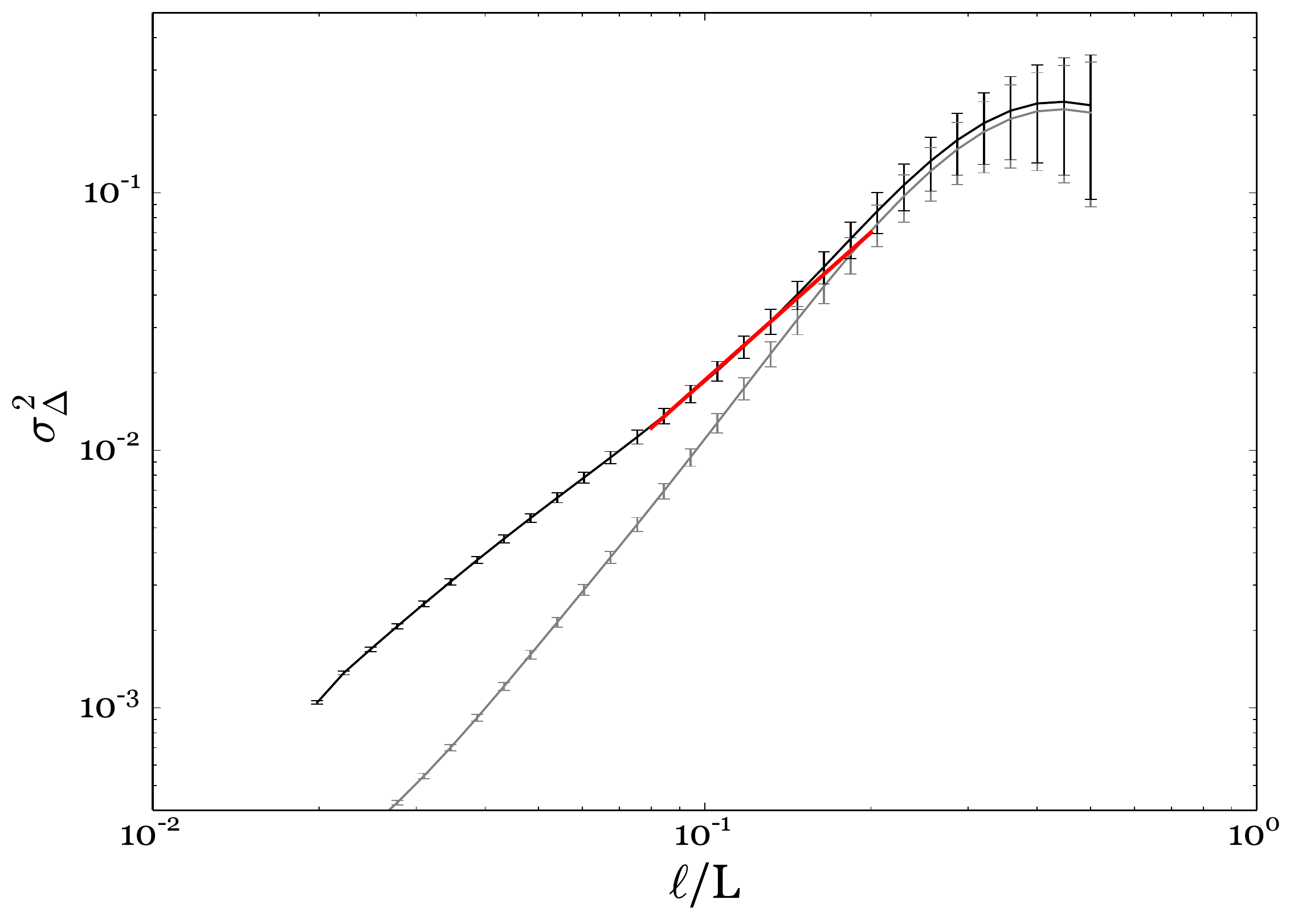}
\caption{The $\Delta$-variance spectrum of G0.253+0.016 as a function of the fractional length scale $\ell /L$, calculated from the column densities derived from the SCUBA-only image (grey line), and from the combined SMA plus SCUBA image (black line). The displayed errors are calculated via the method outlined in \citet{bensch01}. The red line shows the best fit to the combined SMA plus SCUBA $\Delta$-variance between 0.08 and 0.2, giving a slope $1.907\pm0.037$ corresponding to $\alpha=-1.907\pm0.037$. \label{deltavar}}
\end{figure}

We determined the $\Delta$-variance spectrum $\sigma^2_{\Delta}$ \citep{stutzki98} of G0.253+0.016, to compare its observed power-law exponent $\alpha$ to recent predictions linking $\alpha$ to the instantaneous star-formation efficiency \citep[SFE,][]{federrath13}. The power-law exponent of the $\Delta$-variance spectrum determined from the column densities is also the negative exponent of the three-dimensional density power spectrum $\alpha$ \citep{federrath13}, and thus -- as it presents fewer difficulties to determine from observations than power-spectra -- serves as a more robust measure of the cloud structure on a range of scales. Figure \ref{deltavar} presents the $\Delta$-variance spectrum of G0.253+0.016, derived using a Mexican hat kernel, as a function of the fractional size scale $\ell/L$, where $\ell$ is the absolute size scale in the image, and $L$ is the maximum scale in the map, which we took to be the length of the smallest axis. The grey line shows the $\Delta$-variance using column densities determined from only the SCUBA image, and the black line shows the $\Delta$-variance of the column densities derived from the combined SCUBA plus SMA 1.3\,mm continuum image. The error bars shown were calculated as described in \citet{bensch01}. These authors also determined that the $\Delta$-variance spectrum is affected by the observed resolution up to 4$\times$ the beam FWHM. Thus, as $L=272$\,pixels and the beam sizes in pixels are 5.4 and 16 for the combined and SCUBA images respectively, the lower limits for reliable recovery of the $\Delta$-variance are respectively $\ell/L = 0.08$ and 0.24. Thus we fit the combined $\Delta$-variance spectrum between 0.08 and 0.2 to compare it to the results of \citet{federrath13}, who fit their simulations over the range $\ell/L = 0.06$ to 0.2. The best-fitting line with a slope of $1.907\pm0.037$ and thus $\alpha= -1.907\pm0.037$ is shown in red in Figure \ref{deltavar}. As all of the simulations of \citet{federrath13} with $\alpha<-1$ had a SFE$\sim$0\%, we therefore conclude once more that the structure of G0.253+0.016 is consistent with no or little star formation.

The lack of star formation in G0.253+0.016 can also be assessed with reference to proposed critical density thresholds for star formation. For instance, such a threshold for star formation has been proposed by \citet{lada10} to be 0.024\,g\,cm$^{-2}$. However, all of the mass of the cloud we measure, $\sim$10$^{5}$\,M$_{\odot}$ above a limit of 2$\times10^{22}$\,cm$^{-2}$ or 0.094\,g\,cm$^{-2}$, lies above this threshold. Therefore, all $\sim$10$^{5}$\,M$_{\odot}$ should be forming stars, corresponding to 4$\times10^4$ young stellar objects (YSOs), which are not detected \citep{kauffmann13}. These results caution against the idea of one density threshold for star formation, and this can be rationalised on the grounds of what is expected from the interplay of physical conditions and physics to initiate gravitational collapse and hence star formation within the cloud. If we assume that turbulence dominates over magnetic and thermal forces, this interplay is most easily seen in the expression for the virial mass of a cloud, or any core or clump within the cloud,

\begin{equation}
M_{\rm vir} = \frac{5 R \sigma_{\rm v}^2}{G \alpha_{\rm vir}}\,.
\end{equation}

\noindent where $R$ is the radius of the cloud, clump or core, $\sigma_{\rm v}^2$ is the velocity dispersion, $G$ is the gravitational constant, and $\alpha_{\rm vir}$ is the virial parameter \citep[e.g.][]{bertoldi92}, which is of the order of one when the gas is bound. For a bound cloud, clump or core of a given radius, the threshold column density should therefore go as $N_{\rm th} \propto M_{\rm vir}/R^2 \propto \sigma_{\rm v}^2$. 

Similarly, \citet{padoan13} show that the volume density threshold for star formation should depend only on the local linewidth-size relation. This implies that a different threshold is expected for regions with a different linewidth-size relation, such as the Galactic centre \citep[e.g.][]{shetty12}.

Taking the linewidths observed in our combined $^{13}$CO SMA plus IRAM 30m observations over the area covered by G0.253+0.016 (33\,km\,s$^{-1}$, corresponding to $\sigma_{\rm v}=14$\,km\,s$^{-1}$), and comparing it to those measured for nearby clouds \citep[$\sigma_{\rm v}\sim$\,2.5\,km\,s$^{-1}$,][]{kainulainen13}, we would thus expect the threshold density for G0.253+0.016 to scale by a factor of $(14/2.5)^2 = 31.4$ in comparison to the local clouds from which the Lada et al. density threshold was determined. Hence the scaled threshold for star formation in G0.253+0.016 would then be 0.75\,g\,cm$^{-2}$ or $N_{H_2} =$1.6$\times10^{23}$\,cm$^{-2}$. However, the mass above this column density threshold is still 5.4$\times10^{4}\pm^{15}_{13}$\%\,M$_{\odot}$, where the uncertainty is determined by the three assumed values of $\beta$. Using the prescription given by \citet{lada10}, the number of YSOs expected for this mass is 0.18 YSOs/M$_{\odot} \times 5.4 \times10 ^{4}$\,M$_{\odot}$ = 9.7$\times10^{3}$ YSOs. Assuming this number of YSOs, a \citet{kroupa02} IMF with an upper slope of -2.3, and that \citet{lada10} detect all of the YSOs above a limit of 0.08\,M$_{\odot}$, we still expect 10 YSOs of mass $>$15\,M$_{\odot}$ (for the original 4$\times10^{4}$ YSOs we would expect 44). Given a Poisson distribution with an expected value of 10, there is thus a 97\% probability that five or more YSOs would be observed. Hence it is unlikely that we would not observe any YSOs due to small number statistics.

However, we note in Section \ref{SMA_cont} that there are several mm cores with associated with 1.3\,cm free-free continuum emission (Mills et al., in preparation), and there is a water maser observed towards Core 18. Therefore it is possible that 1.3\,cm free-free continuum emission may in fact account for some of the expected YSOs with mass $>$15\,M$_{\odot}$. 

We investigated whether these massive YSOs would be detected at infrared wavelengths. We inserted fake sources into the 70\,$\mu$m \textit{Herschel} PACS image, which contains no detected protostars \citep{longmore12}, and determined that the detection limit was $\sim$10\,Jy due to the high background in the region, which is a factor of $\sim$100 higher than in regions of low background in the outer Galaxy. Given an average column density of $\sim 2 \times10^{23}$cm$^{-2}$ H$_2$ from our analysis of the column density PDF of G0.253+0.016 in Section \ref{PDFs}, and the Milky Way dust properties with $R_V=5.5$ from \citet{draine03a,draine03b}, we determined the optical depth through the cloud at this wavelength to be $\sim$1. Therefore sources should be detected with intrinsic fluxes above 10\,Jy/ $e^{-1}$ = 27 Jy. We then used the models of \citet{robitaille06} to determine the YSO mass required to be detected above a flux limit of 27\,Jy, and found that model YSOs greater than 10\,M$_{\odot}$ should be detectable. Thus, even accounting for the increased level of turbulence in G0.253+0.016 relative to local clouds, there are still several missing massive YSOs that should be detected in the infrared.

As there still remain missing massive YSOs, further aspects may need to be taken into account in the interpretation of these results. Firstly, that the IMF describes the distribution of stars not YSOs. Thus, if there exist massive YSOs in the cloud that are still accreting, they will not have reached their final luminosities (although they should have sizeable accretion luminosities) and may therefore currently be undetectable. Secondly, the lack of observed star formation in G0.253+0.016 could be due to a minimum lag time required for star formation to be observable once a fraction of the gas lies above the threshold for star formation. The detection of water masers in the densest core in G0.253+0.016 suggests an age for the cloud of at least 10$^4$ years \citep{breen10}. This is also in agreement with the results of \citet{federrath13}, which show that the star formation efficiency, although highly dependent on the initial conditions, becomes non-negligible at some fraction of the free-fall time. For a molecular hydrogen density of 10$^4$ to 10$^5$\,cm$^{-3}$, the free fall time is several 10$^5$ years, and thus gives an upper limit for the age of G0.253+0.016, in agreement with the age derived from the water maser detection. Therefore G0.253+0.016 may simply be at an early stage of collapse, and thus evolutionary effects should be taken into account when determining the true star forming efficiency above a given (column) density threshold. In the particular case of clouds in the CMZ, larger-scale dynamics may also be important in understanding their evolution, such as increased cloud collisions or passage close to Sgr\,A* \citep{longmore13}.

In addition, our illustration above does not take into account the background density over which G0.253+0.016 in the CMZ or a cloud in the local ISM is enhanced, which should be removed when determining the boundness of the cloud. The star formation rate in clouds could instead be determined by a critical over-density factor, as put forward or discussed by \citet{krumholz05}, \citet{padoan11a} and \citet{kruijssen14}, above the average density of the cloud or its surrounding medium, depending simply on the energy in turbulence in the cloud, measured by $\sigma_{\rm v}$. \citet{kainulainen14} have also recently shown observationally that there appears to be a constant critical volume over-density for star formation. However, \citet{padoan13} have also argued that the critical volume density may be independent of cloud density (see their equation 10).

In the context of the Kennicutt-Schmidt relation \citep{kennicutt98}, these results suggest that if the observed beam-averaged column or surface densities are dominated by clumps or clouds of similar levels of turbulence $\sigma_{\rm v}$ and average density, the density threshold for star formation will be the same for each observed conglomeration of clouds which provide a point on the relation. As the CMZ only constitutes a small ($\sim$400\,pc) section of the Milky Way, on a size scale only beginning to be resolved for external galaxies, we therefore expect that clouds in the disks of galaxies, which have similar average densities and levels of turbulence to one another, will dominate observations, producing a direct correlation between the star formation rate and the surface density.

\section{Conclusions}
\label{conclusions}

To scrutinise the dynamics and structure, as well as determine the star-forming potential, of the massive Galactic centre cloud G0.253+0.016, we have carried out a concerted SMA and IRAM 30m study of this enigmatic cloud in dust continuum, CO isotopologues, several shock tracing molecules such as CH$_3$OH and SiO, as well as H$_2$CO to trace the gas temperature. In our study, we have also included ancillary far-IR and sub-mm \textit{Herschel} and SCUBA data to further the interpretation of the cloud structure. Our main results are as follows:

\begin{enumerate}

\item We have characterised the 36 cores detected in G0.253+0.016 with the SMA in 1.3\,mm and 1.37\,mm continuum, with masses between 25 and approximately 250\,M$_{\odot}$, comparing these to recent 1.3\,cm VLA observations (Mills et al., in preparation) which suggest that several cores are associated with free-free ionised gas continuum and thus may be tracing the first signs of massive star formation in this cloud.

\item By modelling H$_2$CO line ratios, we find that the kinetic temperature of the gas is extremely large ($>$320\,K) on the size-scales traced by the SMA beam ($\sim$4.3$\times$2.9$''$ or 0.18$\times$0.12\,pc). These temperatures are much hotter than those found for that of the dust, which we find to reach below 20\,K in the innermost regions of the cloud, in agreement with previous results.

\item We observe widespread shock emission over G0.253+0.016, which is strongest in the southern regions of the cloud. Further, by comparing position-velocity diagrams of the large-scale $^{13}$CO emission to that of the shock tracing molecules, we find that G0.253+0.016 intersects in position and velocity with another cloud which peaks at 70\,km\,s$^{-1}$. The shock tracers are brightest and display large velocity gradients close to this interaction zone, indicating cloud-cloud collision. Hence, we have found the first evidence of the specific cloud colliding with G0.253+0.016.

\item By investigating the dynamics of the entire Galactic centre region, we find that the HNC Galactic longitude-velocity diagram of the CMZ is consistent with the CMZ being orientated with Sgr\,B2 on the near-side. We also determine that if this cloud is in fact colliding with another cloud at 70\,km\,s$^{-1}$, a more complex geometry for the CMZ is required than a simple elliptical ring structure.

\item We determined the column density PDF of G0.253+0.016 derived from SMA and SCUBA dust continuum emission is log-normal with no discernible power-law tail, consistent with little star formation. We also find that the width of the column density PDF is narrow but that, given the level of turbulence in the Galactic centre and the enhanced magnetic field in this region, it is consistent with the expected column density PDF created by supersonic, magnetised turbulence.

\item We also investigate the $\Delta$-variance spectrum of this region and show it is consistent with that expected for clouds with 0\% SFE, supporting the fact we see no evidence for star formation from the column density PDF.

\item We show via a simple argument using the virial mass that the star formation column density threshold for G0.253+0.016 should be increased due to the increased turbulence in the CMZ compared to clouds in the Galactic disk, yet incorporating turbulence might still not account for the lack of massive star formation observed. In addition to the level of turbulence, the background or average density may also play a role in the determination of the local column density threshold, making it instead a critical over-density. Thus we confirm that there is no one column density threshold for star formation, but this is dependant on local cloud conditions, and suggest that the Kennicutt-Schmidt relation for external galaxies is observed simply because galactic disk clouds have similar average densities and levels of turbulence, and dominate over clouds in the galactic nucleus such as G0.253+0.016.

\end{enumerate}

\begin{acknowledgements}
We thank the referee, M.G. Burton, and the editor, Malcolm Walmsley, for providing insightful comments which improved this paper. Thank you to Sarah Sadavoy for enlightening discussions on SCUBA data, and to Manuel Gonzalez for helping us with the observations and reductions during a week of bad weather at the IRAM 30m. We are also grateful to Paul Clark, Betsy Mills and Jouni Kainulainen for useful discussions, and to Betsy Mills and collaborators for sending us their 1.3\,cm data before publication. We would like to thank Steve Longmore for providing us with his temperature map for comparison with our results. We are also grateful to Thomas Robitaille with help interpreting his grid of models. 

We used APLpy, an open-source plotting package for Python hosted at http://aplpy.github.com, and Astropy, a community-developed core Python package for Astronomy \citep{astropy13}. This research used the facilities of the Canadian Astronomy Data Centre operated by the National Research Council of Canada with the support of the Canadian Space Agency.

The Mopra CMZ molecular line survey data was obtained using the Mopra radio telescope, a part of the Australia Telescope National Facility which is funded by the Commonwealth of Australia for operation as a National Facility managed by CSIRO. The University of New South Wales (UNSW) digital filter bank (the UNSW-MOPS) used for the observations with Mopra was provided with support from the Australian Research Council (ARC), UNSW, Sydney and Monash Universities, as well as the CSIRO.

This research made use of Montage, funded by the National Aeronautics and Space Administration's Earth Science Technology Office, Computation Technologies Project, under Cooperative Agreement Number NCC5-626 between NASA and the California Institute of Technology. Montage is maintained by the NASA/IPAC Infrared Science Archive.

SER is supported by grant RA2158/1-1 through the Deutsche Forschungsgemeinschaft priority program 1573 (Physics of the Interstellar Medium). 

\end{acknowledgements}

\bibliography{} 

\clearpage
\end{document}